\documentclass[useAMS,usenatbib]{mnras}
\usepackage{amsmath}
\usepackage{graphicx}
\usepackage[utf8]{inputenc}
\usepackage[T1]{fontenc}
\usepackage{aecompl}
\title
[\rm \batman]
{\batman: Bayesian Technique for Multi-image Analysis}
\author[J.~Casado et al.]
{
J.~Casado$^{1,2,3}$\thanks{E-mail:javier.casado@uam.es},
Y.~Ascasibar$^{1,2}$,
R.~Garc\'{i}a-Benito$^{4}$,
G.~Guidi$^{3}$,
\\~\\
{\normalfont\LARGE
O.~S.~Choudhury$^{3}$,
E.~Bellocchi$^{1,2}$,
S.~F.~S\'anchez$^{5}$
and A.I. D\'{i}az$^{1,2}$
}
\\~\\
$^{1}$ Universidad Aut\'onoma de Madrid, 28049 Madrid, Spain\\
$^{2}$ Astro-UAM, UAM, Unidad Asociada CSIC\\
$^{3}$ Leibniz-Intitut f{\"u}r Astrophysik Potsdam (AIP), D-14482, Potsdam, Germany\\
$^{4}$ Instituto de Astrof\'{i}sica de Andaluc\'{i}a (IAA/CSIC), 18080 Granada, Spain\\
$^{5}$Instituto de Astronom\'{i}a, Universidad Nacional Aut\'onoma de M\'exico, A.P. 70-264, 04519, M\'exico
}

\date{\bf Accepted (\today)}

\pagerange{\pageref{firstpage}--\pageref{lastpage}}
\pubyear{2016}

\usepackage[usenames,dvipsnames,svgnames,table]{xcolor}

\newcommand{\dd}{{\rm d}}
\newcommand{\xx}{\ensuremath{x_{ij\lambda}}}
\newcommand{\ee}{\ensuremath{e_{ij\lambda}}}
\newcommand{\rr}{\ensuremath{r_{ij}}}
\newcommand{\zz}{\ensuremath{\theta_{r\lambda}}}
\newcommand{\mm}{\ensuremath{\mu_{r\lambda}}}
\newcommand{\sg}{\ensuremath{\sigma_{r\lambda}}}
\newcommand{\signal}{\ensuremath{S_{ij\lambda}}}
\newcommand{\residual}{\ensuremath{\Delta_{ij\lambda}}}
\newcommand{\residualB}{\ensuremath{\hat\Delta_{ij\lambda}}}
\newcommand{\nrow}{\ensuremath{n_{\rm row}}}
\newcommand{\ncol}{\ensuremath{n_{\rm col}}}
\newcommand{\nreg}{\ensuremath{n_{\rm reg}}}

\newcommand{\likelihood}{\ensuremath{\mathcal{L}(X|E,R,\Theta)}}
\newcommand{\prior}{\ensuremath{\mathcal{P}}}

\newcommand{\figs}{figs}

\newcommand{\ngc}{NGC\,2906}
\newcommand{\ha}{H$\alpha$}
\newcommand{\hb}{H$\beta$}
\newcommand{\hg}{H$\gamma$}
\newcommand{\hd}{H$\delta$}

\newcommand{\batman}{{\sc BaTMAn}}
\newcommand{\shifu}{{\sc shifu}}

\usepackage{hyperref}
\begin{document}

\maketitle

\label{firstpage}

%--------------------------------------------------------------------------
\begin{abstract}
This paper describes the Bayesian Technique for Multi-image Analysis (\batman), a novel image-segmentation technique based on Bayesian statistics that characterizes any astronomical dataset containing spatial information and performs a tessellation based on the measurements and errors provided as input.
The algorithm iteratively merges spatial elements as long as they are statistically consistent with carrying the same information (i.e. identical signal within the errors).
We illustrate its operation and performance with a set of test cases including both synthetic and real Integral-Field Spectroscopic data.
The output segmentations adapt to the underlying spatial structure, regardless of its morphology and/or the statistical properties of the noise.
The quality of the recovered signal represents an improvement with respect to the input, especially in regions with low signal-to-noise ratio.
However, the algorithm may be sensitive to small-scale random fluctuations, and its performance in presence of spatial gradients is limited.
Due to these effects, errors may be underestimated by as much as a factor of two.
Our analysis reveals that the algorithm prioritizes conservation of all the statistically-significant information over noise reduction, and that the precise choice of the input data has a crucial impact on the results.
Hence, the philosophy of \batman\ is not to be used as a `black box' to improve the signal-to-noise ratio, but as a new approach to characterize spatially-resolved data prior to its analysis.
The source code is publicly available at \url{http://astro.ft.uam.es/SELGIFS/BaTMAn}.
\end{abstract}
%--------------------------------------------------------------------------


%--------------------------------------------------------------------------
\section{Introduction}
\label{sec_introduction}
%-------------------------------------------------------------------------

One of the basic problems in astronomical data analysis is the characterization of spatially-resolved information and the measurement of physical properties (and their variation) across extended sources.
Soon this kind of tasks became increasingly demanding as new observations started to provide larger and larger amounts of data, prompting the need for a certain level of automation.
Nowadays, human supervision is often limited to problematic cases, and many current and forthcoming datasets are so vast that a significant part of the analysis is left entirely to computer programs.

One of the first instances of such application is the identification of (potentially extended and/or blended) sources in photometric images.
Since the advent of large extragalactic surveys in the late 70's, a wide variety of techniques have been developed in order to automatically create source catalogues from astronomical images \citep[e.g.][among many others]{Stetson87, Bertin&Arnouts96, Makovoz&Marleau05, Savage&Oliver07, Molinari+11, Hancock+12, Menshchikov+12}.
The widespread use of Integral-Field Spectroscopy (IFS) has literally added a new dimension to the problem, in the sense that spatial and spectral information are combined in order to locate the sources \citep[see e.g.][and references therein, in the context of radio observations of the 21-cm line]{Koribalski12}.

Recently, new algorithms working on different types of spatially-resolved data have been developed in order to tackle specific scientific problems beyond source detection.
In many cases, the aim is to compute and characterize maps that trace the spatial distribution of a given physical quantity
% other than the surface brightness at any given wavelength,
such as e.g. the temperature and composition of the hot intracluster medium \citep{Sanders&Fabian01, Diehl&Statler06}, the properties of the stellar population in early-type galaxies \citep{Cappellari&Copin03}, the moments of the velocity distribution along the line of sight \citep{Krajnovic+06}, or the emission of warm ionized ionized gas and {\sc Hii} regions in galaxies \citep{Papaderos+02, HIIexplorer, Pipe3D}.

What most of these goals have in common is the challenge of performing a coherent spatial segmentation of an image (or an IFS datacube) as a first step.
On the one hand, it is necessary to average the signal over a large region in order to increase the signal-to-noise ratio (hereafter S/N) and carry out meaningful measurements.
On the other hand, averaging over too large area does not only lead to a loss in spatial resolution, but it may also (in some cases, strongly) bias the results and their physical interpretation if the signal within the chosen region is not homogeneous.

As mentioned above, this problem has already been tackled by different authors. They have proposed several schemes, specifically optimized for a wide variety of problems, to divide an image into connected regions in a fully automatic way.
Some of them \citep[e.g.][]{Sanders&Fabian01, Cappellari&Copin03, Diehl&Statler06} are based on a Voronoi tessellation, whereas some others rely upon the identification of suitable intensity thresholds \citep[e.g.][]{Stetson87, Bertin&Arnouts96, HIIexplorer} or isocontours \citep[e.g.][]{Papaderos+02, Sanders06} at specific wavelengths in order to define physically-motivated regions.
More recently, \citet{Pipe3D} proposed an alternative method, also aiming to obtain a tessellation with a target S/N, that imposes `continuity' in the surface brightness (i.e. a maximum contrast within any region) in order to better adapt to the morphology of the data.

This work takes another step in this direction, presenting an alternative approach that is not aimed to obtain a specific S/N but to identify spatial regions where the signal is statistically consistent with being constant within the provided errors: if two regions carry the same information, it will always be advantageous to merge them in order to (further) increase the signal-to-noise ratio; if they are `different' (do not carry the same information), they should be kept separate in order not to introduce artificial biases.
Such prescription preserves the information contained in the input dataset, and it imposes no condition on the shape of the tessellation.
It is extremely general, and it can be applied to any kind of spatially-resolved data.

We describe the mathematical basis of the method in Section~\ref{sec_idea}, while the details of the algorithm and its implementation are discussed in Section~\ref{sec_code}.
Section~\ref{sec_data} proposes a set of benchmark problems, and the analysis of the results is discussed in Section~\ref{sec_results}.
Our main conclusions are summarized in Section~\ref{sec_conclusions}.

%-------------------------------------------------------------------------
\section{Mathematical formulation of the problem}
\label{sec_idea}
%-------------------------------------------------------------------------

%------------------------------------------------
\begin{table*}
\begin{center}
\begin{tabular}{lllc}
\hline
             & Symbol                & Meaning & Equation \\
\hline
\hline
DIMENSIONS   & $i,j$                 & spatial dimensions (bi-dimensional, $\nrow\times\ncol$ maps) \\
             & $\lambda$             & spectral dimension ($n_\lambda$ `layers' or `wavelengths') \\
\hline 
INPUT        & $D=\{X,E\}$           & input data set (measurements and errors) given to \batman & \eqref{eq_data} \\
             & $X = \{ \xx \}$       & measurements at every 3D position \\
             & $E = \{ \ee \}$       & errors associated to every 3D position \\
\hline
MODEL        & $M=\{ R, \Theta \}$   & assumed model (tessellation and signals) to describe the data & \eqref{eq_model} \\
             & $R = \{ \rr \}$       & tessellation (labels associated to every spatial position) \\
             & $\Theta = \{ \zz \}$  & signal associated to every region $r$ and wavelength $\lambda$ \\
             & $\likelihood$         & likelihood of the measurements, given the errors and the model & \eqref{eq_likelihood} \\
\hline
PRIORS       & $\prior(R,\Theta) = \prior(R)\,\prior(\Theta|R)$
                                     & prior probability distribution of the model parameters \\
             & $\prior(\Theta|R)$    & prior probability of the signals $\Theta$ for a given tessellation $R$ & \eqref{eq_prior_theta} \\
             & $\prior(R)$           & prior probability of the tessellations $R$ & \eqref{eq_prior_R} \\
\hline
POSTERIORS   & $p(R,\Theta|D)$  & posterior probability of the model parameters, given the data & \eqref{eq_posterior} \\
             & $p(\Theta|R,D)$  & posterior probability of the signals $\Theta$, given $R$ and $D$ & \eqref{eq_posterior_Theta}, \eqref{eq_posterior_Theta_Gaussian} \\
             & $p(R|D)$         & posterior probability of the tessellations $R$, given $D$ & \eqref{eq_posterior_seg} \\
             & \mm              & expected value of the signal according to $p(\Theta|R,D)$ & \eqref{eq_mean} \\
             & \sg              & expected variance (`formal error') of the expected signal & \eqref{eq_variance} \\
\hline
EVIDENCE    & $\mathcal{E}$         & overall Bayesian evidence for our assumed description & \eqref{eq_evidence}, \eqref{eq_evidence_2} \\
             & $\mathcal{E}(R)$      & evidence for the tessellation $R$ & \eqref{eq_evidence_R1}, \eqref{eq_evidence_R_full} \\
\hline
NUMERICAL      & $K$                 & combined prior parameter & \eqref{eq_K} \\
IMPLEMENTATION & $R_{n}$             & currently-accepted tessellation with $n$ regions \\
               & $R_{c,n-1}$         & candidate tessellations with $n-1$ regions & \eqref{eq_posterior_candidate} \\
\hline
TEST CASES   & $S_{ij\lambda}$       & signal associated to every 3D position (Eq.~\ref{eq_test_cases}) \\
AND          & $e_{ij\lambda}$       & noise associated to every 3D position (Eq.~\ref{eq_test_cases}) \\
ANALYSIS     & $e_{{\rm ccd}}$, $e_{{\rm uniform}}$   &  different error prescriptions, see Section~\ref{sec_synthetic} \\
             & \residual             & residual shown in Figure~\ref{fig_error} (Eq.~\ref{eq_residual1})   \\
             & \residualB            & residual shown in Figure~\ref{fig_sigma} (Eq.~\ref{eq_residual1})  \\
\hline
\end{tabular}
\end{center}
\caption
{
List of symbols used in the article.
}
\label{tab_symbols}
\end{table*}
%------------------------------------------------

\batman\ (Bayesian Technique for Multi-image Analysis) is a new binning scheme designed to perform an adaptive segmentation on a three-dimensional dataset that serves as inceptive step for its further analysis.
We will refer to these sets as `datacubes' or `multi-images' along the manuscript, as they consist of an arbitrary number ($n_\lambda$ `layers' or `wavelengths') of spatially-arranged two-dimensional data (individual `images' or `maps').
We will focus on IFS data throughout the paper, and we will use the term `spaxel' to denote the individual and minimum spatial element of a dataset.
However, the procedure can be applied to any set of images, or to any kind of spatially-distributed data.
In particular, the algorithm can also be applied to a single image (i.e. single-`layer', 2D dataset) in what we will refer to as `monochromatic' binning mode.

More precisely, \batman\ works on an \emph{input dataset} $D=\{X,E\}$ consisting of two $\nrow \times \ncol$ multi-images $X$ and $E$, with $N=\nrow \times \ncol$ spaxels each, whose spatial positions we will tag with the indices $\{i,j\}$.
Every spaxel has a set of $n_\lambda$ `measurements' $\xx$ and `errors' \ee
\begin{equation}
\begin{split}
& X = \{ \xx \}\\
& E = \{ \ee \}
\end{split}
\label{eq_data}
\end{equation}
associated to its spatial location, where $i=[1,\nrow]$, $j=[1,\ncol]$, and $\lambda=[1,n_\lambda]$.
Different `layers' may correspond to broad-band observations of the same object, different wavelengths in an IFS datacube, or even maps with completely different information (e.g. mass surface density, velocity, velocity dispersion, age, metallicity, equivalent width, etc).

The goal of the algorithm is to divide the input dataset into spatially-connected `regions' that carry the same information, i.e. where all the measurements \xx\ are statistically compatible with being random samples of the same underlying signal \signal\ given the Gaussian errors \ee.
In the current implementation, \batman\ makes use of Bayesian statistics to connect adjacent spaxels into `regions' (and merge adjacent regions into larger ones) as long as it is more likely that the true values of the measured quantities (the `signal') are constant within the merged region ($\signal=\zz\ \forall\{i,j\}$ in region $r$) than a collection of several independent domains with different values.
More elaborate (e.g. quadratic) models of the underlying signal may be considered in future versions of the code.

Thus, the \emph{output dataset} is an optimized `segmentation model' $M=\{ R, \Theta \}$ consisting of $\nreg$ connected regions (with every spaxel $\{i,j\}$ assigned to the region \rr) where the signal at the different bands is assumed to be well described by the constant values \zz:
\begin{equation}
\begin{split}
& R = \{ \rr \}\\
& \Theta = \{ \zz \}
\end{split}
\label{eq_model}
\end{equation}
with $r=[1,\nreg]$.
$R$ refers thus to the tessellation of the image, fully specified by a $\nrow \times \ncol$ matrix containing the labels \rr\ assigned to every spaxel, and $\Theta$ corresponds to the set of $\nreg \times n_\lambda$ parameters \zz\ used to describe the signal.

According to the frequentist approach, the best segmentation model would be given by the values of $R$ and $\Theta$ that maximise the \emph{likelihood}, i.e. the probability of obtaining the measurements $X$, given the errors $E$ and the $\nreg$ regions specified by $R$ and $\Theta$.
The model proposed for this work assumes that the observed values \xx\ are statistically-independent random variables, with Gaussian probability distributions whose dispersions are given by the errors \ee\ while their means $\theta_{r_{ij}\lambda}$ are to be determined.
With these considerations, one can express the likelihood as
\begin{equation}
\begin{split}
\likelihood
&=
\prod_{i=1}^{\nrow} \prod_{j=1}^{\ncol} \prod_{\lambda=1}^{n_\lambda}
\frac{ e^{ -\frac{(\xx-\theta_{r_{ij}\lambda})^2}{2\ee^2} } }
     { \sqrt{2\pi\ee^2} }\\
&=
\prod_{r=1}^{\nreg} \prod_{i,j\in r} \prod_{\lambda=1}^{n_\lambda}
\frac{ e^{ -\frac{(\xx-\zz)^2}{2\ee^2} } }
     { \sqrt{2\pi\ee^2} }
\end{split}
\label{eq_likelihood}
\end{equation}
where the second product in the last term simply loops over the spaxels in a given region.

Within the Bayesian framework \batman\ is based upon, the approach is slightly different, and it comprises several additional ingredients.
One must declare not only a likelihood for the data given the model, but also a \emph{prior} probability distribution $\prior(R,\Theta)$ of the model parameters.
The construction of this prior is based on our previous knowledge about the problem, and it will certainly affect the inferences we obtain from our data.
As an example, it could favour certain tessellations \citep[roundish bin shapes, such as in][]{Cappellari&Copin03, HIIexplorer, Pipe3D} or avoid/penalize unphysical values (e.g. negative values of the signal).

The Bayesian approach combines the prior with the likelihood of observing the measurements $X$ (making use of Bayes' theorem) to compute the \emph{posterior} probability distribution of the model parameters
\begin{equation}
p(R, \Theta | D) = \frac{ \prior(R,\Theta)\ \likelihood }{ \mathcal{E} }
\label{eq_posterior}
\end{equation}
given the data.
The overall Bayesian \emph{evidence} for our assumed description
\begin{equation}
\mathcal{E} \equiv p(X|E) = \sum_{R} \int \prior(R,\Theta)\ \likelihood\ \dd\Theta
\label{eq_evidence}
\end{equation}
is the sum over all possible tessellations $R$ and signal values $\Theta$, and it can be considered as the overall probability of observing the measurements $X$, given the errors $E$, if our model provided an accurate description of the data (in frequentist terms, the likelihood of our description).
The expected values of the model parameters and their uncertainties can be computed from the posterior probability distribution~\eqref{eq_posterior}.

In \batman, the problem of image segmentation is split in two parts: 
\begin{enumerate}
 \item the estimation of the signal distribution $\Theta$ given a fixed segmentation $R$
 \item the further selection of the most probable tessellation
\end{enumerate}

The exact procedures followed to solve every one of them are discussed in the following sections. 

\subsection{Parameter estimation}
\label{sec_param}
%-------------------------------------------------------------------------

Given a fixed tessellation $R$, \batman\ will evaluate the posterior probability distribution $p(\Theta|R,D)$ describing the signal \zz\ in every region $r$ at every wavelength $\lambda$.

Following a Bayesian procedure, we first assume that the measurements $X$ are described by the Gaussian likelihood function (see Eq.~\ref{eq_likelihood}).
Then, as we do not expect any specific distribution for the values of $\zz \in \Theta$, we have adopted a uniform (sometimes referred to as `noninformative' or `objective') prior.
The uniform probability distribution assigns equal probability to all values of the parameter, reflecting our initial ignorance.
In order to obtain a \emph{proper} prior, for which the total probability is normalized to unity (in contrast to an \emph{improper} prior, where the integral diverges to infinity), the possible values of the signal \zz\ are restricted to a finite range of values:
\begin{equation}
\begin{split}
 \prior(\zz|R) & ~\dd\zz = \frac{1}{H_\lambda-L_\lambda}~\dd\zz \\
 & L_\lambda < \zz < H_\lambda
\end{split}
\label{eq_prior_theta_k1}
\end{equation}
where upper and lower limits $H_\lambda$ and $L_\lambda$ correspond to the highest and lowest values measured at that particular wavelength, anywhere within the image (i.e. the maximum and minimum \xx\ over $i$ and $j$ for a given $\lambda$).

From a practical point of view, this prescription automatically selects a reasonable range for the signal (irrespective of e.g. measurement units in the different layers of the multi-image), 
but, from a strict Bayesian perspective, one may object that we have actually looked at the data in order to set our prior.
In order to generalise expression~\eqref{eq_prior_theta_k1}, we introduce $n_\lambda$ additional parameters $0<k_\lambda\le 1$ such that
\begin{equation}
\begin{split}
\prior(\Theta|R) \ \dd \Theta
& = \prod_{\lambda=1}^{n_\lambda} \prod_{r=1}^{\nreg} \prior(\zz|R)~\dd\zz \\
& = \prod_{\lambda=1}^{n_\lambda} \left( \frac{ k_\lambda }{ H_\lambda-L_\lambda } \right)^{\!\!\nreg} \dd\zz \\
L_\lambda - \Delta_\lambda &< \zz < H_\lambda + \Delta_\lambda
\end{split}
\label{eq_prior_theta}
\end{equation}
where $\Delta_\lambda = \frac{1-k_\lambda}{2k_\lambda}(H_\lambda-L_\lambda)$.
The choice $k_\lambda=1$ results in equation~\eqref{eq_prior_theta_k1}, whereas $k_\lambda=0$ would yield the improper prior $-\infty<\zz<\infty$.
Thus, the value of $k_\lambda$ provides a qualitative indication of the weight given to the information contained in $H_\lambda$ and $L_\lambda$.

Using Bayes' theorem, the posterior probability distribution for the signal is given by
\begin{equation}
p(\Theta | R, D)~\dd\Theta
= \frac{ \prior(\Theta|R)~\likelihood }
       { \mathcal{E}(R) }~\dd\Theta
\label{eq_posterior_Theta}
\end{equation}
where \likelihood\ is the likelihood defined in equation~\eqref{eq_likelihood}, and the evidence for the tessellation $R$ can be expressed as
\begin{equation}
 \mathcal{E}(R) = p(X | E,R) = \int \prior(\Theta|R)~\likelihood ~\dd \Theta \\
 \label{eq_evidence_R1}
\end{equation}

Since we have adopted a uniform prior, and our model likelihood is a product of $\nrow\times \ncol\times n_\lambda$ Gaussians, it is easy to verify\footnote{Using that
$ \sum_i \frac{(x-\mu_i)^2}{2\sigma_i^2} = \frac{(x-\mu)^2}{2\sigma^2} - \frac{\mu^2}{2\sigma^2} + \sum_i \frac{\mu_i^2}{2\sigma_i^2} $ with $\frac{1}{\sigma^2}\equiv\sum_i\frac{1}{\sigma_i^2}$ and $\mu\equiv\sigma^2\sum_i\frac{\mu_i}{\sigma_i^2}$.} that the posterior probabilities
\begin{equation}
p(\zz | R, D) \propto e^{ -\frac{(\zz-\mm)^2}{2\sg^2} }
\label{eq_posterior_Theta_Gaussian}
\end{equation}
are independent Gaussians with mean
\begin{equation}
\mm = \sg^2 \sum_{i,j \in r} \frac{\xx}{\ee^2}
\label{eq_mean}
\end{equation}
and dispersion
\begin{equation}
\frac{1}{\sg^2} = \sum_{i,j \in r} \frac{1}{\ee^2}
\label{eq_variance}
\end{equation}
for any value of the prior parameters $k_\lambda$.

In other words, the expected value \mm\ of the signal within any given region $r$ reported by \batman\ (and the associated `formal errors' \sg) are obtained from an \emph{inverse-variance weighted average} over the region.

Finally, the evidence for the tessellation $R$ is given by
\begin{equation}
\begin{split}
\mathcal{E}(R)
&=
\int
\prod_{r=1}^{\nreg}
\prod_{\lambda=1}^{n_\lambda}
\prior(\zz|R)
\prod_{i,j\in r}
\frac{ e^{ -\frac{(\xx-\zz)^2}{2\ee^2} } }
     { \sqrt{2\pi\ee^2} }
\, \dd\zz\\
&=
\prod_{r=1}^{\nreg}
\prod_{\lambda=1}^{n_\lambda}
\frac{ k_\lambda }
     { H_\lambda-L_\lambda }
\frac{ \int e^{ -\sum_{ij\in r} \frac{(\xx-\zz)^2}{2\ee^2} } \, \dd\zz }
     { \prod_{ij\in r} \sqrt{2\pi\ee^2} }
\\
&\equiv
\prod_{r=1}^{\nreg} \mathcal{E}_r
\end{split}
\label{eq_evidence_R_full}
\end{equation}
with
\begin{equation}
\begin{split}
\mathcal{E}_r
&\equiv
\prod_{\lambda=1}^{n_\lambda}
\frac{ k_\lambda\ \int\! e^{ -\frac{(\zz-\mm)^2}{2\sg^2} + \frac{\mm^2}{2\sg^2} - \sum_{ij\in r} \frac{\xx^2}{2\ee^2} } \, \dd\zz }
     { (H_\lambda-L_\lambda)\ \prod_{ij\in r} \sqrt{2\pi\ee^2} }\\
&=
\prod_{\lambda=1}^{n_\lambda}
\frac{ k_\lambda\ e^{ \frac{\mm^2}{2\sg^2} - \sum_{ij\in r} \frac{\xx^2}{2\ee^2} }\ \sqrt{2\pi\sg^2} }
     { (H_\lambda-L_\lambda)\ \prod_{ij\in r} \sqrt{2\pi\ee^2} }
\end{split}
\label{eq_evidence_region}
\end{equation}

\subsection{Model selection}
%-------------------------------------------------------------------------

Once we obtain the evidence $\mathcal{E}(R)$ for any tessellation $R$ and the posterior probability distribution of the recovered signal $\Theta$ associated to $R$, we face the question of selecting the `optimal' segmentation that is most likely to describe our dataset.
This is the second time we apply a Bayesian approach to our dataset, but two main differences exist with respect to the parameter estimation problem described in the previous section:
\begin{enumerate}
 \item the set of all possible tessellations is discrete, whereas the values of the signal are continuous variables
 \item we will require \batman\ to select one \emph{and only one} `optimal' tessellation rather than a probabilistic description (which is, as we will discuss later, not `strictly Bayesian').
\end{enumerate}

Except for these two differences, the Bayesian approach to model selection is very similar to parameter estimation: first, we must specify our priors $\prior(R)$ and then combine them with the appropriate likelihood $p(X|E,R)$ in order to obtain the posterior probability distribution $p(R|D)$ from Bayes' theorem.

We have no preliminary information relative to the preferred tessellations, and \batman\ will not impose any geometrical constraint (e.g. roundness) other than ensuring that all the regions defined by the matrix \rr\ are physically connected.
% 
% In principle we have little preliminary information relative to the preferred tessellations.
% \batman\ will ensure that all the regions defined by the matrix \rr\ are physically connected, but no other geometrical constraint (e.g. roundness) is imposed.
It is however foreseen that some practical applications of the algorithm may have preference for a larger or smaller number of regions $\nreg$, and therefore we have chosen a prior of the form
\begin{equation}
\prior(R) = k_{\rm R}^{ \nreg }
\label{eq_prior_R}
\end{equation}
where $k_{\rm R}$ is to be defined by the user.
The value $k_{\rm R}=1$ corresponds to the improper uniform prior, that assigns equal probability to all valid segmentations regardless of the number of regions they contain; $k_{\rm R}<1$ would favour that the multi-image is divided into a small number of (large) regions, whereas $k_{\rm R}>1$ would favour that individual spaxels are kept independent or cluster into a large number of small regions.

The likelihood of a tessellation $R$ (i.e. the probability of measuring $X$, considering all possible values of $\Theta$) is simply the evidence $\mathcal{E}(R)$ defined in equation~\eqref{eq_evidence_R1}, and thus the posterior probability distribution of $R$ is given by
\begin{equation}
 p(R|D) =  \frac{ \prior(R)\, \mathcal{E}(R) }{ \mathcal{E} }
 \label{eq_posterior_seg}
\end{equation}
where $\mathcal{E}$ denotes the overall evidence~\eqref{eq_evidence}.
Due to the discrete nature of $R$,
\begin{equation}
 \mathcal{E} =  \sum_R \prior(R)~\mathcal{E}(R)
 \label{eq_evidence_2}
\end{equation}
where the $R$ runs over all possible tessellations.

The current version of \batman\ outputs the matrix \rr\ (labels corresponding to the tessellation) that maximises the posterior probability $p(R|D)$, along with the corresponding values of \mm\ and \sg.
Let us argue at this point that, to some extent, the very concept of `model selection' is intrinsically not Bayesian.
If, for example, $p(R_+|D)=0.5+\delta$ and $p(R_-|D)=0.5-\delta$, one can only choose $R_+$, no matter how small $\delta$ may be, if forced to take a decision.
We do think that a strictly Bayesian algorithm would never select a unique segmentation, but calculate the posterior probability of all possible tessellations and weight them in order to provide a fully probabilistic description of the underlying signal, spaxel by spaxel.
However, such approach would yield a `smoothing' technique rather than a `segmentation' tool, and it will not be discussed further in the present work.

%-------------------------------------------------------------------------
\section{Numerical implementation}
\label{sec_code}
%-------------------------------------------------------------------------

\batman\ is based on the philosophy that the purpose of a segmentation algorithm is to group together the `spaxels' of a `multi-image' that carry the same information (measurements \xx\ compatible with identical signal \signal\ within the errors \ee) into larger `regions'.
According to the arguments presented in Section~\ref{sec_idea}, this goal can be mathematically formulated as the maximisation of the Bayesian posterior probability of the tessellation $R$, fully described by the $\nrow \times \ncol$ matrix \rr\ specifying the label of the region that each spaxel $(i,j)$ belongs to.
Once $R$ is fixed, we have shown that the posterior probability distribution of the `signal' \zz\ within each region $r$ at `wavelength' (or `layer') $\lambda$ is Gaussian, with mean \mm\ and standard deviation \sg\ given by an inverse-variance weighted average over the region, i.e. equations~\eqref{eq_mean} and~\eqref{eq_variance}, respectively.

In practice, the number of possible tessellations is so large that evaluating all their evidence is completely infeasible.
Therefore, \batman\ follows a greedy iterative procedure, merging adjacent regions until no further increase in the posterior probability $p(R|D)$ is possible:

\begin{enumerate}

\item A unique label $l_{ij}=[1, N]$ is assigned to each spaxel.
The initial segmentation $R_N$ considers that every spaxel is an independent region, i.e. $\nreg=N$, $\rr=l_{ij}$, $\mm=\xx$, and $\sg=\ee$.
From now on, we will drop the subscript and denote the number of regions simply as $n$.

\item On every iteration, starting from $n=N$, \batman\ compares the posterior probability of $R_n$ with all possible candidate tessellations $R_{c,n-1}$ that can be obtained by merging any two adjacent\footnote{We consider contiguous spaxels in the horizontal or vertical directions, but not along the diagonals.} regions in $R_n$.
More precisely, it evaluates the ratios $\frac{ p(R_{c,n-1}|D) }{ p(R_n|D) }$ for all these candidate tessellations with $n-1$ regions each, where the subscript $c$ runs over all the candidates considered.

\item The algorithm selects the optimal tessellation for the next iteration, $R_{n-1}$, as the candidate displaying the highest probability (i.e. the largest ratio with respect to $R_n$).
If none of the $R_{c,n-1}$ is found to be more likely than $R_n$, no further iterations will be performed.
The greedy procedure is illustrated in Appendix~\ref{sec_1D_problem}, and the exit condition is discussed in Appendix~\ref{sec_stopping}.

\item On exit, \batman\ outputs the matrices containing the labels \rr\ of the different regions, the mean posterior signal maps \mm, and and their standard deviation \sg\ for the final `optimized' tessellation $R_n$.

\end{enumerate}

A critical part of the algorithm is thus the evaluation of the probability ratios
\begin{equation}
\begin{split}
\frac{ p(R_{c,n-1}|D) }{ p(R_n|D) }
&= \frac{ \prior(R_{c,n-1})\, \mathcal{E}(R_{c,n-1}) }{ \prior(R_n)\, \mathcal{E}(R_n) } \\
&= \frac{ \prior(R_{c,n-1})\ \prod_{r'\in R_{c,n-1}} \mathcal{E}_{r'} }{ \prior(R_n)\ \prod_{r\in R_n} \mathcal{E}_r }
\end{split}
% \label{eq_ratio_posterior}
\end{equation}
where $r$ and $r'$ refer to the regions defined in $R_n$ and $R_{c,n-1}$, respectively.
Due to our iterative procedure, these two models only differ in that two regions of $R_n$ (let us name them $A$ and $B$) that have been merged into a single region $A \cup B$ in $R_{c,n-1}$.
Substituting expression~\eqref{eq_evidence_region} for the contribution of each region to the evidence and equation~\eqref{eq_prior_R} for the prior probability of each tessellation, one arrives to
\begin{equation}
\begin{split}
\frac{ p(R_{c,n-1}|D) }{ p(R_n|D) }
&=
\frac{ k_{\rm R}^{n-1} }{ k_{\rm R}^n~ } \frac{ \mathcal{E}_{A\cup B} }{ \mathcal{E}_A\, \mathcal{E}_B }\\
&=
k_{\rm R}^{-1}
\prod_{\lambda=1}^{n_\lambda}
\frac{ (H_\lambda-L_\lambda)\,\sigma_{A\cup B,\lambda}\ e^{ \frac{\mu_{A\cup B,\lambda}^2}{2\sigma_{A\cup B,\lambda}^2} } }
     { k_\lambda\, \sqrt{2\pi}\,\sigma_{A\lambda}\,\sigma_{B\lambda}\ e^{ \frac{\mu_{A\lambda}^2}{2\sigma_{A\lambda}^2} + \frac{\mu_{B\lambda}^2}{2\sigma_{B\lambda}^2} } }\\
&\equiv
K^{-1}
\prod_{\lambda=1}^{n_\lambda}
(H_\lambda-L_\lambda)
\frac{ e^{-\frac{ (\mu_{A\lambda}-\mu_{B\lambda})^2 }{ 2(\sigma_{A\lambda}^2+\sigma_{B\lambda}^2) }} }
     { \sqrt{2\pi(\sigma_{A\lambda}^2+\sigma_{B\lambda}^2)} }\\
\end{split}
\label{eq_posterior_candidate}
\end{equation}
where the prior parameters $k_{\rm R}$ and $k_\lambda$ can be neatly separated from the term comparing the measurements in regions $A$ and $B$ in terms of their dispersions.
In fact, it is evident from expression~\eqref{eq_posterior_candidate} that the combined effect of $k_{\rm R}$ and $k_\lambda$ can be grouped into a single number
\begin{equation}
 K \equiv k_{\rm R} \prod_{\lambda=1}^{n_\lambda} k_\lambda
\label{eq_K}
\end{equation}
that encapsulates all our choices about the prior probability distributions for \rr\ and \zz.
Since all other quantities (including $H_\lambda$ and $L_\lambda$, which denote the maximum and minimum values of \xx, respectively) are driven by the data, $K$ is the only free parameter of our algorithm.

%-------------------------------------------------------------------------
\section{Test cases}
\label{sec_data}
%-------------------------------------------------------------------------

In order to illustrate the capabilities of the algorithm, assess its performance, and provide general guidelines to understand its operation, we have considered a set of four different input datasets as benchmark problems.
Two of them are based on a synthetic multi-image (Section~\ref{sec_synthetic}) that poses several potentially challenging situations (e.g. different shapes, with and without sharp boundaries, different noise statistics and signal-to-noise ratios, overlap between independent structures, etc.), and the other two are based on integral-field spectroscopic observations of the local galaxy \ngc\ (Section~\ref{sec_obs}).

These four datasets consist of several `layers' or `wavelentghs' (i.e. they are all `multi-images' according to our terminology).
For every one of them we have run \batman\ in two different ways: one considering all the layers simultaneously and obtaining a single common tessellation (we will refer to this approach as `multi-$\lambda$') and another in which we consider every `layer' individually and obtain an independent segmentation for every one them (`monochromatic' approach).
This yields a total of eight test cases, as summarized in Table~\ref{tab_tests}.

For all of them, we carry out two independent runs adopting $K=1$ and $K=10^{-6}$.
The former (obtained e.g. by setting both $k_{\rm R}$ and every $k_\lambda$ to unity) corresponds to a binning criterion that tends to keep small regions separated; reducing the value of $K$ (e.g. by decreasing either $k_{\rm R}$ and/or the product of the different $k_\lambda$) favours the coaddition of spaxels and leads to the definition of larger regions.

\subsection{Synthetic data}
\label{sec_synthetic}
% -------------------------------------------------------------------------

Our synthetic multi-images consist of 3 layers (arbitrarily labelled R, G, and B) that display different objects over a null background where the signal value is set to $S_0=(0,0,0)$.
The R layer shows a quarter of a circle in the bottom-left corner with signal $S_{\rm circle}=(1,0,0)$, the G layer displays a triangle with $S_{\rm triangle}=(0,7,0)$ located the centre of the image, slightly overlapping with the circle, and the B image shows an ellipse centred in the top-right corner (again, slightly overlapping with the triangle) that displays a continuous, radially-decreasing gradient $S_{\rm ellipse}=(0,0,[17-1])$.

This three-layer datacube
\begin{equation}
 S = (S_{\rm R},S_{\rm G},S_{\rm B}) = S_0 + S_{\rm circle} + S_{\rm triangle} + S_{\rm ellipse}
\label{eq_test_signal}
\end{equation}
represents the \emph{signal} of our synthetic problem.
From these data, we generate two different \emph{input} multi-images
\begin{equation}
 \xx = S_{ij\lambda} + e_{ij\lambda}
 \label{eq_test_cases}
\end{equation}
by adding random Gaussian noise $n$ with different statistics.
The `uniform-noise' datacube follows a Gaussian probability distribution with zero mean and dispersion $e_{\rm uniform}=(1,1,1)$ for all layers of every spaxel. We also generate a `CCD-like' datacube (mimicking the characteristics of charge-coupled devices) as a result of adding a white-noise component with dispersion $\sigma_{\rm W}^2=(0.5,0.5,0.5)$ and a Poisson noise component where the dispersion $\sigma_{\rm P}^2$ grows linearly with the signal $S_{ij\lambda}$ at any given band and location, according to the expression
\begin{equation}
e_{{\rm ccd}:ij\lambda}^2 = \sigma_{\rm W}^2 + \sigma_{{\rm P}:ij\lambda}^2
= \frac{ 1 + S_{ij\lambda} }{ 2 }
\label{eq_ccd_noise}
\end{equation}

The numeric values of the intensities and noise levels in our synthetic test problems have been chosen so that the signal-to-noise ratio
\begin{equation}
  (S/N)_{ij\lambda} = S_{ij\lambda} / \ee
\end{equation}
covers a range from one to $17$ in the `uniform' case and up to approximately $6$ for the `CCD-like' noise.
The precise values for the errors and signal-to-noise ratios within each region are summarized in Table~\ref{tab_sn}.
 
%------------------------------------------------
\begin{table}
\begin{center}
\begin{tabular}{lccc}
\hline
 & $S$ & $e_{\rm ccd}$ & $(S/N)_{\rm ccd}$ \\
\hline
Background  &    $0$   &    $0.5$     &  $0$          \\
R: Circle   &    $1$   &    $1$      &  $1$     \\
G: Triangle &    $7$   &    $2$  &     $3.5$ \\
B: Ellipse  & $[1-17]$ & $[1-3]$ & $[1-5.67]$ \\
\hline
\end{tabular}
\end{center}
\caption
{
Values of the signal $S$ of each individual component of the synthetic test multi-image~\eqref{eq_test_signal} in the relevant layer (R,G, or B), followed by the corresponding `CCD-like' noise $e_{\rm ccd}$ given by expression~\eqref{eq_ccd_noise} and the signal-to-noise ratio $(S/N)_{\rm ccd}=S/e_{\rm ccd}$.
For the `uniform-noise' datacube, $e_{\rm uniform}=(1,1,1)$ implies $(S/N)_{\rm uniform}=S/e_{\rm uniform}=S$.
}
\label{tab_sn}
\end{table}
%------------------------------------------------

%------------------------------------------------
\begin{table*}
\begin{center}
\begin{tabular}{cccccc}
\hline
    & TEST CASE       & INPUT: ($\nrow \times \ncol \times n_\lambda$) & $n_{\rm output}$ & FIGURES & PANELS \\
\hline
\#1 & synthetic + uniform noise      & R: ($73 \times 78 \times 1$)    & 3 & 1, 3, 4 & top-right \\
    & `monochromatic'                & G: ($73 \times 78 \times 1$) \\
    &                                & B: ($73 \times 78 \times 1$) \\
\hline
\#2 & synthetic + uniform noise      & ~~ ~($73 \times 78 \times 3$)   & 1 & 1, 3, 4 & top-middle \\
    & `multi-$\lambda$' \\
\hline
\#3 & synthetic + ccd-like noise     & R: ($73 \times 78 \times 1$)    & 3 & 1, 3, 4 & bottom-right \\
    & `monochromatic'                & G: ($73 \times 78 \times 1$) \\
    &                                & B: ($73 \times 78 \times 1$) \\
\hline
\#4 & synthetic + ccd-like noise     & ~~ ~($73 \times 78 \times 3$)   & 1 & 1, 3, 4 & bottom-middle \\
    & `multi-$\lambda$' \\
\hline
\#5 & \ngc: \shifu\ maps          & \ha: ($71 \times 78 \times 1$)  & 4 & 2, 5 & top-right \\
    & `monochromatic'                & \hb: ($71 \times 78 \times 1$) \\
    &                                & \hg: ($71 \times 78 \times 1$) \\
    &                                & \hd: ($71 \times 78 \times 1$) \\
\hline
\#6 & \ngc: \shifu\ maps          & ~~ ~~($71 \times 78 \times 4$)  & 1 & 2, 5 & top-middle \\
    & `multi-$\lambda$' \\
\hline
\#7 & \ngc: CALIFA data           & \ha: ($71 \times 78 \times 14$) & 4 & 2, 5 & bottom-right \\
    & `$\Delta \lambda = \pm 15$\AA' & \hb: ($71 \times 78 \times 15$) \\
    &                                & \hg: ($71 \times 78 \times 15$) \\
    &                                & \hd: ($71 \times 78 \times 15$) \\
\hline
\#8 & \ngc: CALIFA data           & ~~ ~~($71 \times 78 \times 59$) & 1 & 2, 5 & bottom-middle \\
    & `full set' \\
\hline
\end{tabular}
\end{center}
\caption
{
Summary of the test cases presented in Section~\ref{sec_data} and discussed in Section~\ref{sec_results}.
The labelling and nomenclature used to refer each case are quoted on the first two columns, respectively.
The dimensionality of the input data is listed on the third column, whereas the fourth column provides the number of output tessellations (i.e. the number of times \batman\ is run), and the last two columns indicate the location of the corresponding results in the article figures.
Every test case has been run with two different values for the prior parameter, $K=1$ and $K=10^{-6}$.
}
\label{tab_tests}
\end{table*}
%------------------------------------------------

\subsection{Astronomical observations}
\label{sec_obs}
% -------------------------------------------------------------------------

In order to illustrate the use of the algorithm in a typical science case, we consider two different segmentation problems related to the measurement of the intensity of the Balmer emission lines in the local spiral galaxy \ngc.
Observations have been retrieved from the public database of the Calar Alto Legacy Intergral-Field spectroscopic Area (CALIFA) survey \citep{Sanchez+12}.
More precisely, they correspond to the COMBO datacubes delivered in Data Release 2 \citep[see][for a detailed description]{Garcia-Benito+15}, with a uniform wavelength coverage from 3650 to 7200~\AA, sampled in constant steps of 2~\AA.

We will consider two completely different approaches to address the problem.
In the first one, we measure the intensity of the \ha(6563\AA), \hb(4861\AA), \hg(4340\AA), and \hd(4101\AA) emission lines spaxel by spaxel, and then take the resulting intensity maps, with their corresponding errors, as a four-layer input multi-image (i.e. first measure and then bin with \batman).
As an alternative approach, we run \batman\ directly on the raw IFS data, selecting four wavelength intervals of $\pm 15$~\AA\ around each line (see Table~\ref{tab_tests}), and then measure their intensity from the integrated spectrum within each region (i.e. first bin the datacube directly with \batman\ and then measure).

In both cases, we follow exactly the same procedure to measure the intensity of the Balmer lines in a given spectrum.
In order to account for stellar absorption lines, we use {\sc starlight} \citep{Cid-Fernandes+05} to model the spectral energy distribution of the underlying stellar population, fitting the observed spectrum with a linear combination of simple stellar populations spanning different ages and metallicities.
We use the the spectra provided by \citet{Vazdekis+10} for populations older that 64~Myr and \citet{Gonzalez-Delgado+05} models for younger ages.
Dust effects are modelled as a foreground screen, assuming a \citet{Cardelli+89} reddening law with $R_V = 3.1$.
Then, we subtract the estimated stellar continuum from the observed spectrum and obtain the line flux with the SHerpa IFU line fitting software (\shifu; Garc\'{i}a-Benito, in preparation), based on CIAO’s {\sc Sherpa} package \citep{Freeman+01, Doe+07}.
Small deviations with respect to the stellar continuum are taken into account by a first order polynomial, and independent Gaussians have been fitted for the emission lines.

The intensity distribution of the Balmer emission lines in \ngc\ is fairly clumpy (see Figure~\ref{fig_test_galaxy}), with a relatively large number of individual compact {\sc Hii} regions arranged in a ring-like structure.
This configuration is clearly visible in the \ha\ maps, where the signal is stronger, but it is much more difficult to identify in the weakest lines, especially \hd.
In addition, there is a weak diffuse component of much lower surface brightness, arising from a combination of intrinsic emission from diffuse ionized gas \citep{Haffner+09} and light from the compact {\sc Hii} regions that is scattered towards the observer by dust particles in the interstellar medium \citep[see e.g.][]{Ascasibar+16}.

Although the actual solution is obviously not available in this case, all our emission lines arise from the same element, and therefore they should all roughly trace the same spatial distribution (mostly determined by the electron number density, with a secondary dependence on the local electron temperature).
However, the intensity of the lines decreases with frequency, and therefore they probe very different signal-to-noise ratios, from values of the order unity, especially in the outer parts, to well above one hundred in the regions of brightest \ha\ emission.

\subsection{Analysis procedure}
\label{sec_procedure}
% -------------------------------------------------------------------------

As mentioned above, we have approached our four input datasets (both synthetic and astronomical) in two different ways.
On the one hand, we make use of the default binning mode (`multi-$\lambda$') where all the information is considered at the same time (all `layers' binned simultaneously) and one optimal tessellation is computed for the whole dataset.
On the other hand, an alternative `monochromatic' procedure is also considered, where \batman\ is run individually on every layer of the multi-image dataset (hence, several times) and an independent tessellation is derived for every one of them.
A summary of all the resulting test cases is provided in Table~\ref{tab_tests}.
Every test has been repeated twice, adopting the values $K=1$ and $K=10^{-6}$ for the combined prior parameter.

In the synthetic multi-images, \batman\ has been applied to the three layers (R, G, and B) separately (`monochromatic', cases \#1 and \#3) and simultaneously (`multi-$\lambda$', cases \#2 and \#4).
The two test cases based on \shifu\ measurements of \ngc\ (cases \#5 and \#6) follow an approach completely analogous to the synthetic multi-images. The only difference is that now the data consist of four different (but physically not independent) `layers', one for each Balmer line. In case \#5, \batman\ is run in `monochromatic' mode, producing 4 different segmentations, while in case \#6 it is run only once, yielding a single `multi-$\lambda$' tessellation.
For the remaining observational test cases, \#7 and \#8, we have applied \batman\ directly on the IFS data from the CALIFA survey (spectra and error).
We have run our algorithm separately on every 30~\AA-wide slice around each line, thus obtaining 4 independent tessellations (test case \#7), as well as to the full 59-layer multi-image (the four 30~\AA-wide slices together, case \#8), where a single tessellation is returned. Although these tests are scientifically identical to test cases \#5 and \#6 (4 balmer lines binned separately and simultaneously), they are technically different: \batman\ is applied in `multi-image' mode every time, many more `wavelengths' are being considered, and there is no pre-processing of the data (lines are measured with \shifu\ after the binning).

As a final remark, let us note that there is another important difference between test cases \#7 and \#8 (direct application of the algorithm on the IFS datacube) and the rest.
In test cases $\#1-6$, the input data are maps of the measured quantities. \batman\ provides not only an optimal tessellation, but also the expected value \mm\ and the dispersion \sg\ of the signal within every region, based on the posterior probability distribution.
On the contrary, test cases \#7 and \#8 bin directly a selected section of the spectral energy distributions. Although \batman\ returns again the three products (\rr, \mm\ and \sg) only the tessellation is further used by \shifu\ in order to derive the flux of each line and the corresponding error, taking into account noise covariance in CALIFA data \citep[see e.g.][]{Husemann+13, Garcia-Benito+15} as part of its pipeline.
Therefore, the estimated errors may be considerably larger (by about a factor of 3, depending on the number of spaxels within each region) than those implied by equation~\eqref{eq_variance}.

%-------------------------------------------------------------------------
\section{Results}
\label{sec_results}
%-------------------------------------------------------------------------

Based on the test cases described in the previous section, we now address the quality of the output tessellation in terms of its ability to adapt to the underlying signal, the dependence of the results on the adopted priors, the accuracy of the formal errors estimated by the algorithm, and the improvement with respect to the original input data.

%------------------------------------------------
\begin{figure*}
\centering
\includegraphics[height=.4\textheight]{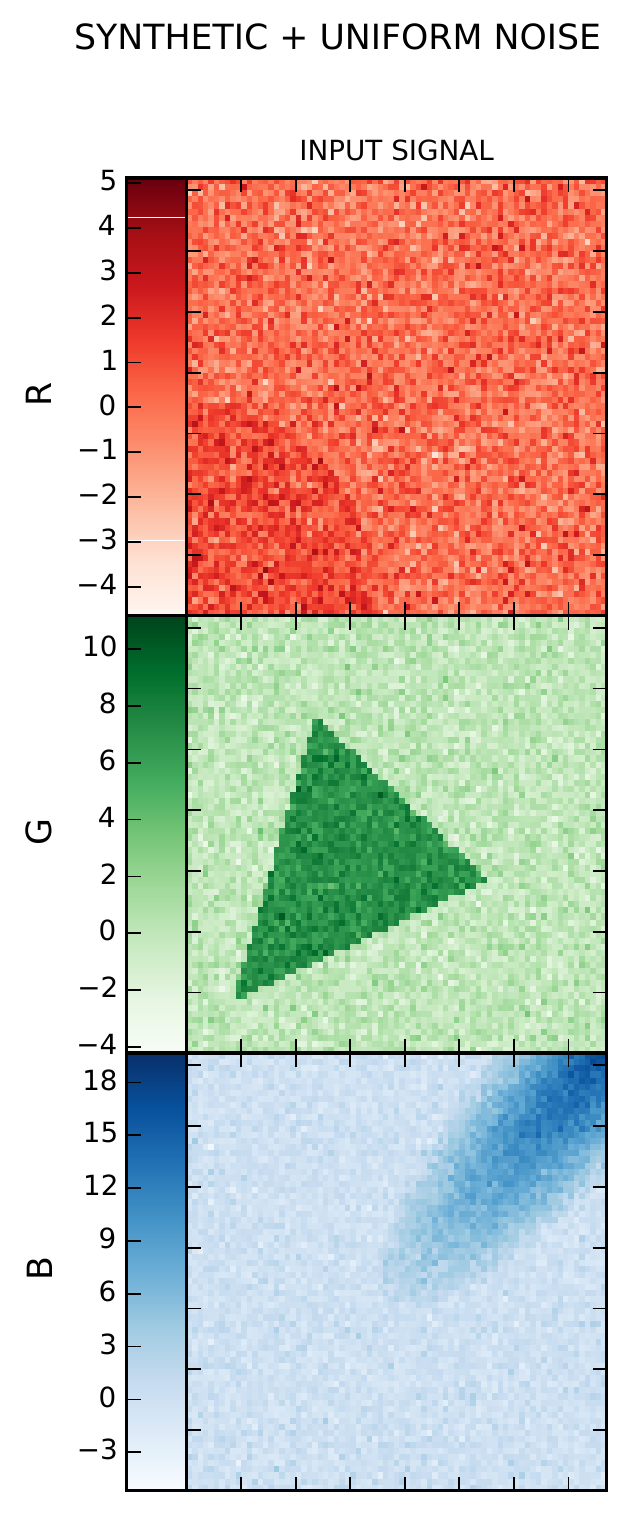} ~~ ~~ ~~
\includegraphics[height=.4\textheight]{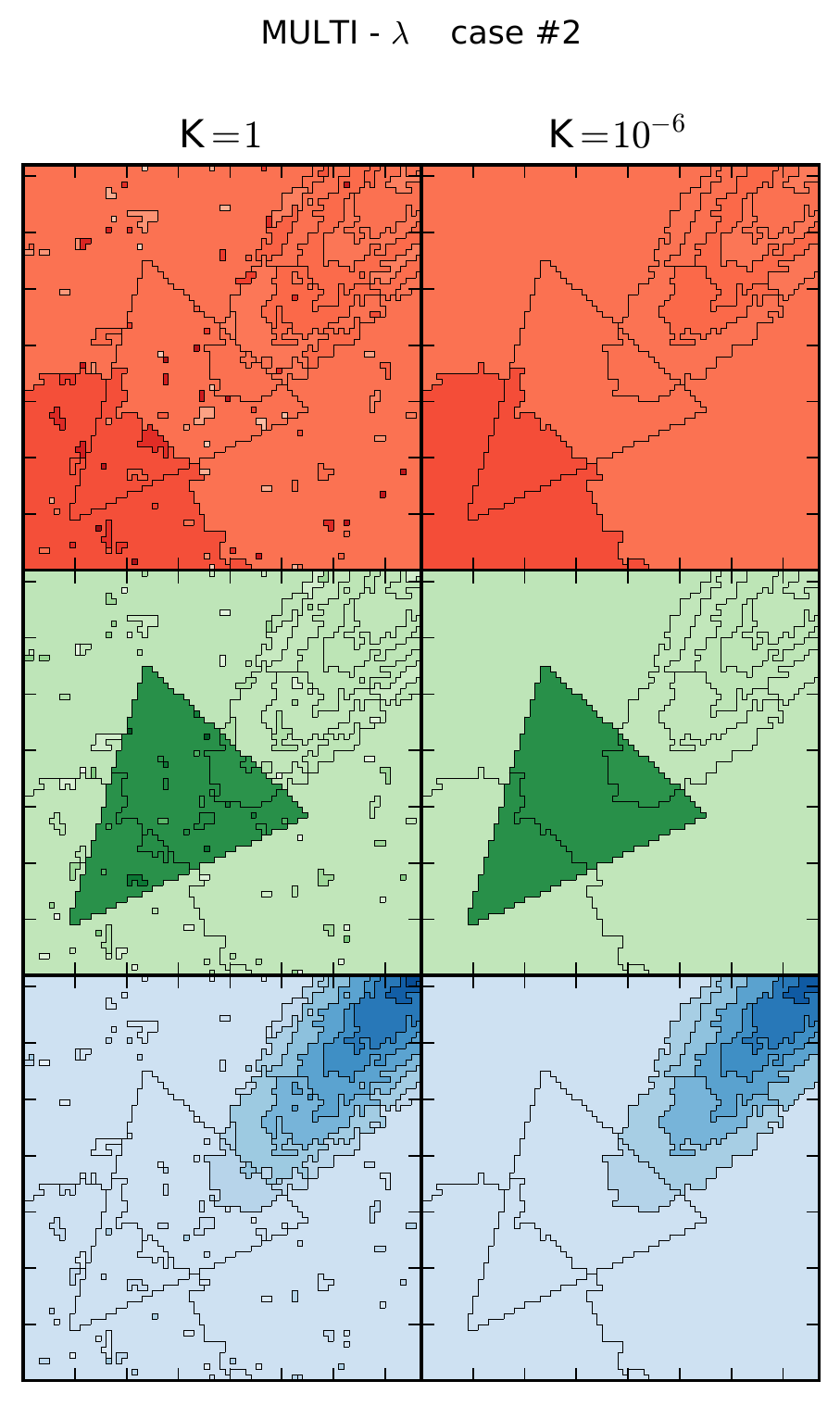}~~
\includegraphics[height=.4\textheight]{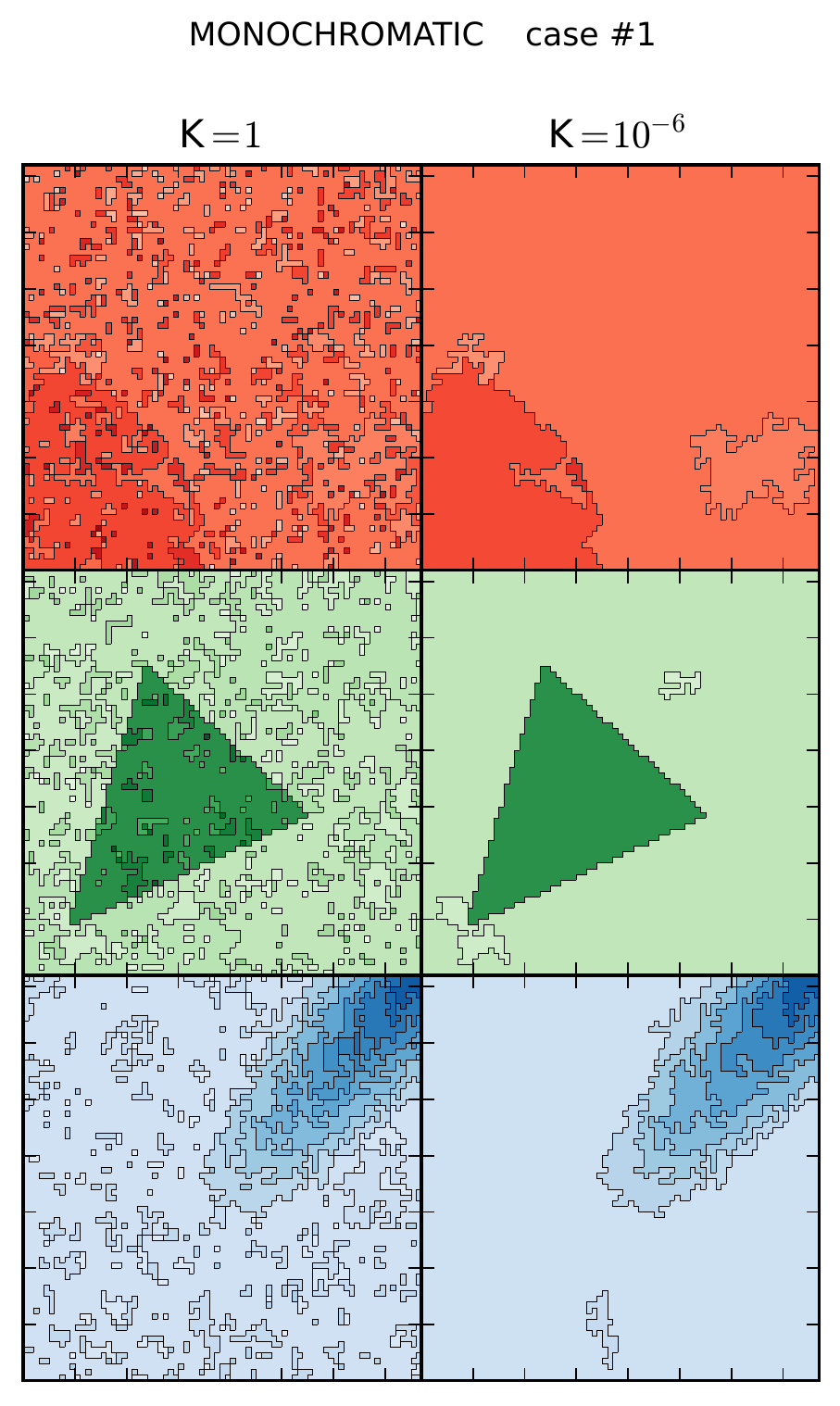}\\[3mm]
\includegraphics[height=.4\textheight]{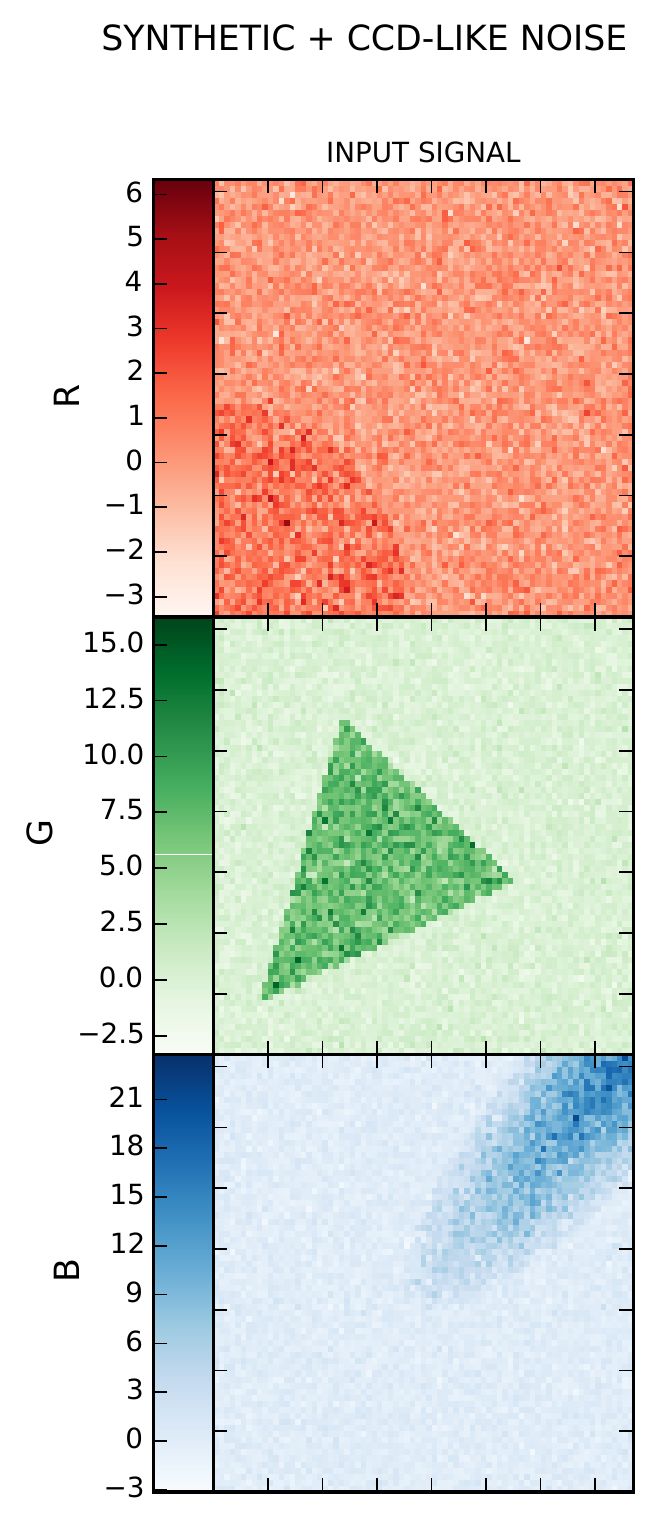} ~~ ~~ ~~
\includegraphics[height=.4\textheight]{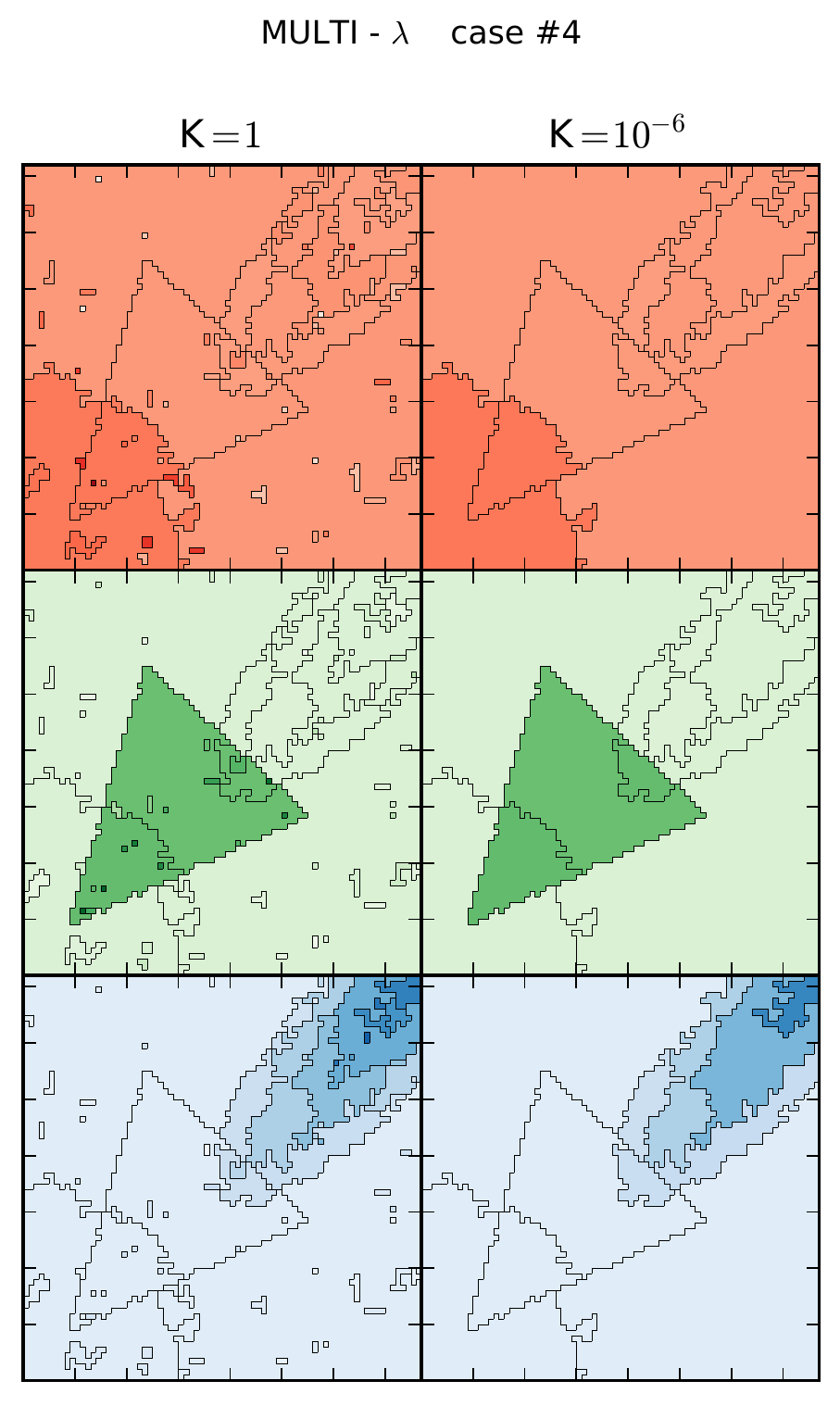}~~
\includegraphics[height=.4\textheight]{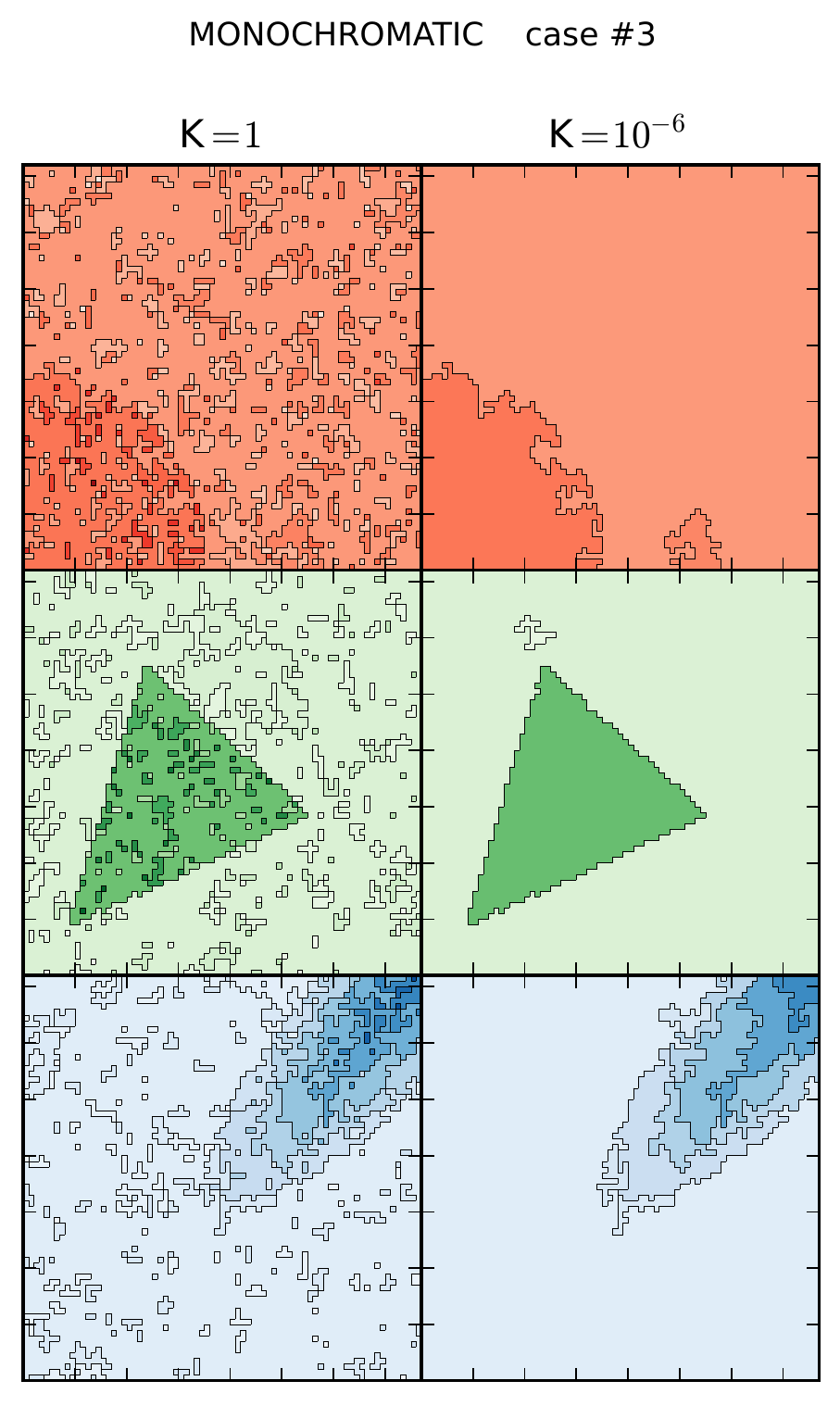}
\caption
{
Results of applying \batman\ to our synthetic test cases in `multi-$\lambda$' mode (i.e. considering all the layers in the input data simultaneously; test cases \#2 and \#4, central columns) and `monochromatic' mode (considering each layer separately, cases \#1 and \#3, right columns). Two runs are performed in every case, with $K=1$ (left) and $K=10^{-6}$ (right). The input signal is displayed in the left column. Rows within each panel correspond to the different layers in the multi-image. The segmentation obtained in every case is indicated with solid black lines and is common to the three layers in the `multi-$\lambda$' test cases \#2 and \#4 and specific to every one in the `monochromatic' test cases \#1 and \#3.
% 
% Results of applying \batman\ with $K=1$ and $K=10^{-6}$ .
% Uniform-noise and CCD-like datacubes are shown on the top and bottom panels, respectively.
% Rows within each panel correspond to the different layers in the multi-image; the input data are displayed in the first column; the two central columns (labelled `multi-$\lambda$') shows the segmentation and the recovered signal when \batman\ is applied to all layers simultaneously, and the binning is thus identical for all of them; the last two (`monochromatic') columns display the results when every layer is considered separately.
}
\label{fig_test_RGB}
\end{figure*}
%------------------------------------------------
\begin{figure*}
\centering
\hfill
\includegraphics[height=.43\textheight]{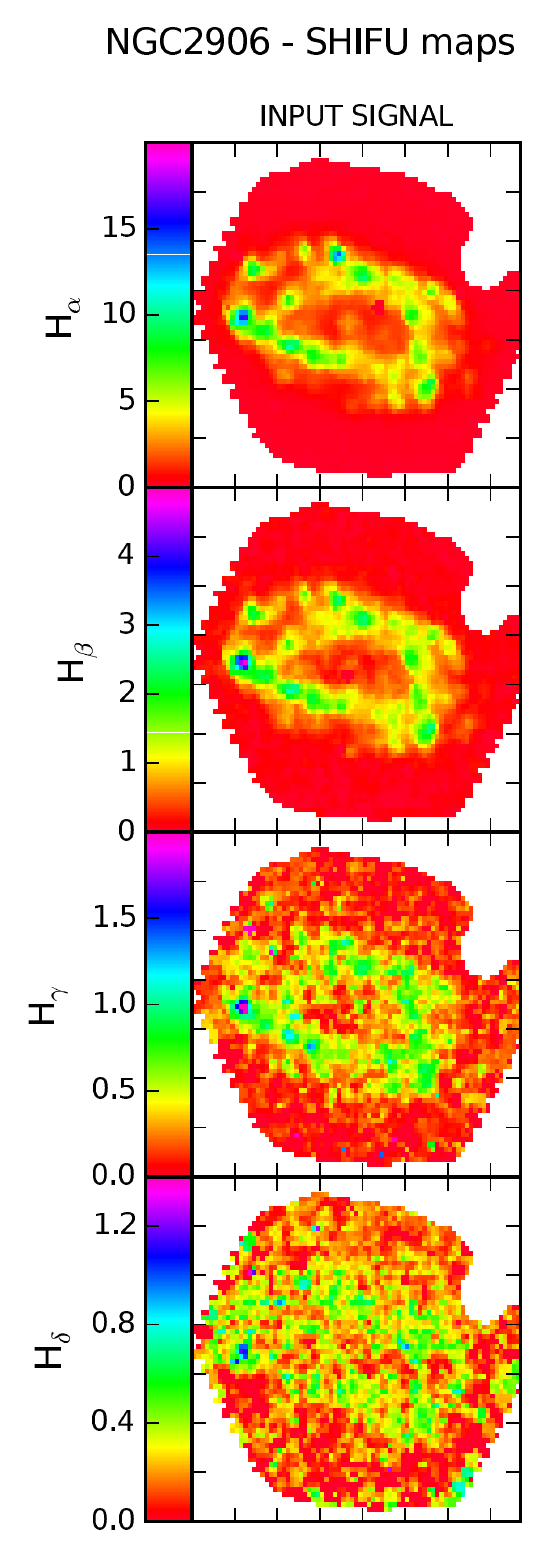}~~
\hfill
\includegraphics[height=.43\textheight]{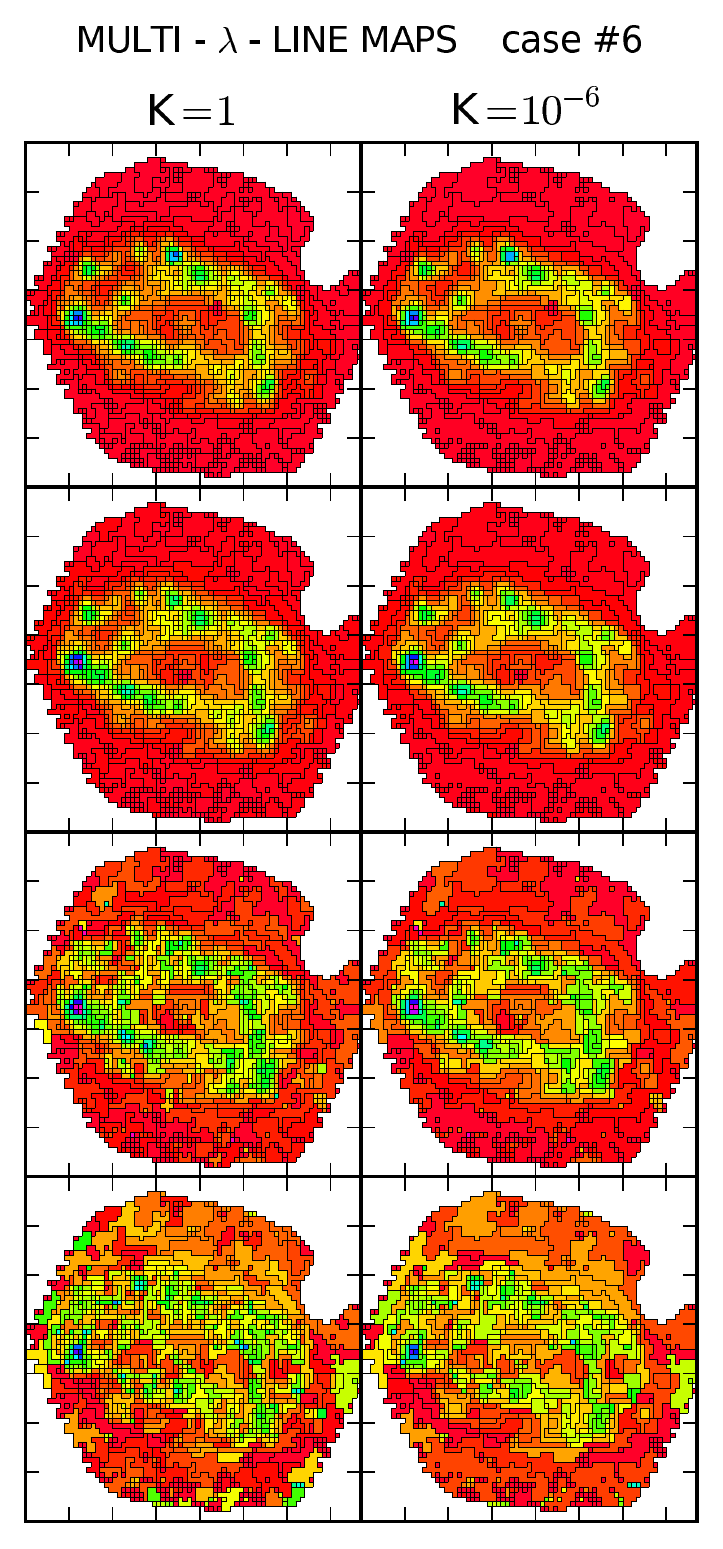}~~
\includegraphics[height=.43\textheight]{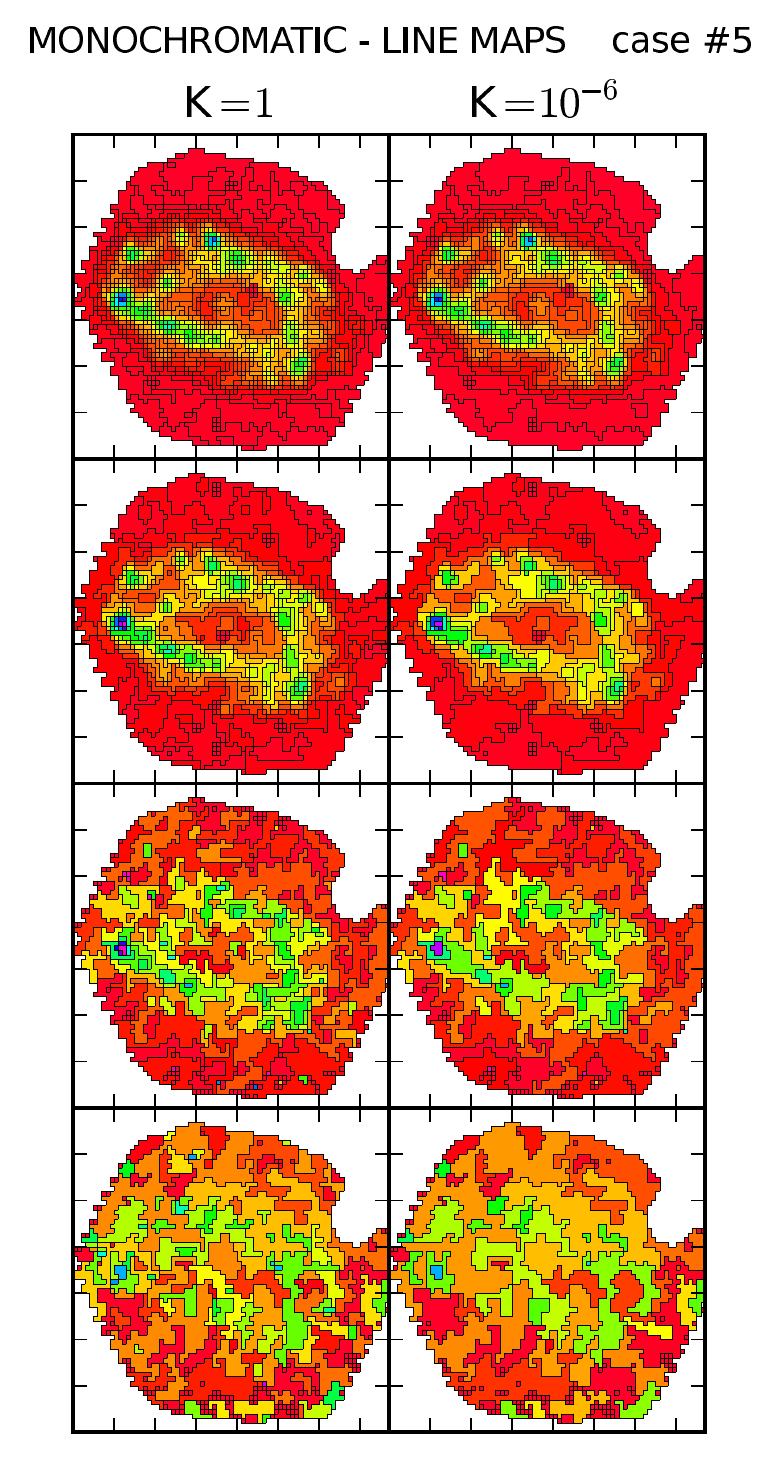}
~~ ~~ ~~ ~~\\[3mm]
% \hfill
\includegraphics[height=.43\textheight]{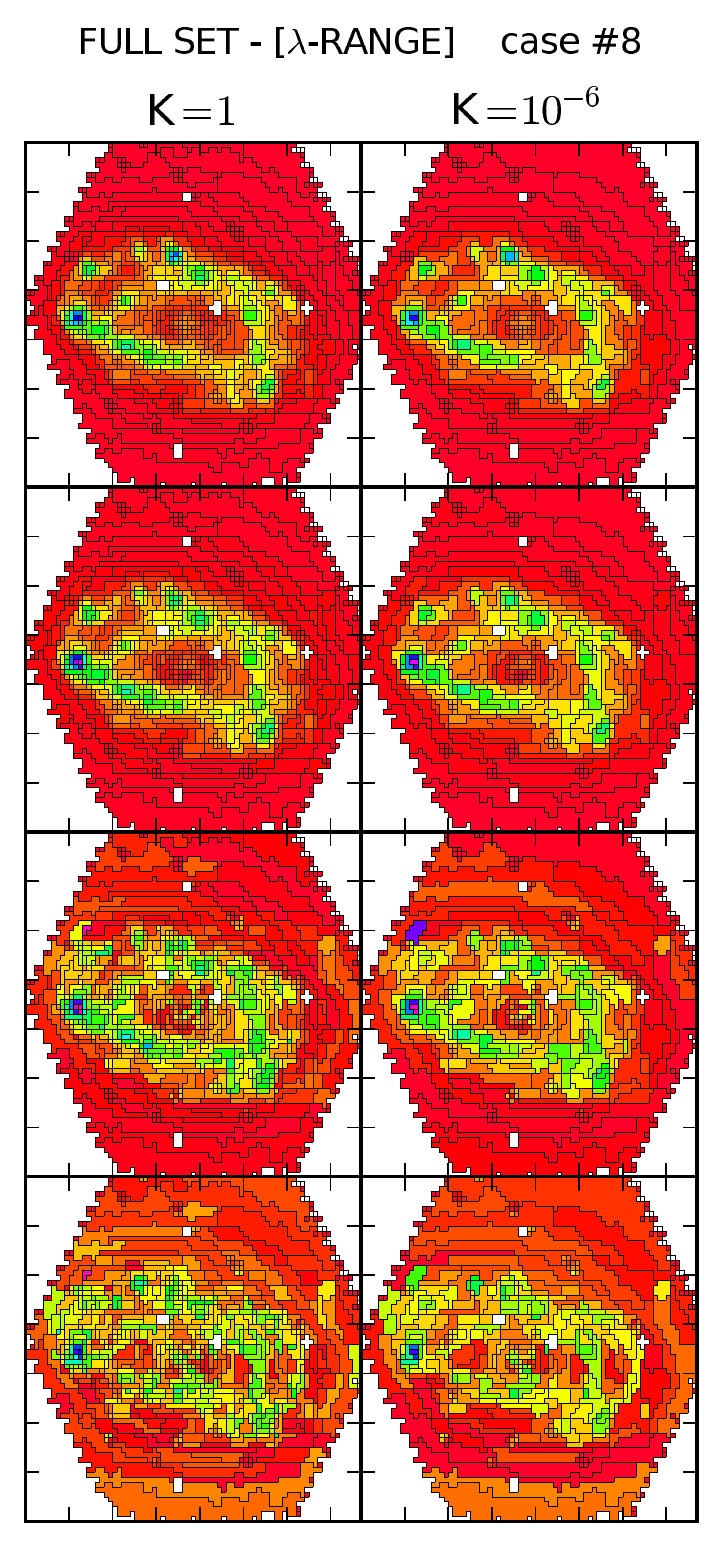}~~ ~~
\includegraphics[height=.43\textheight]{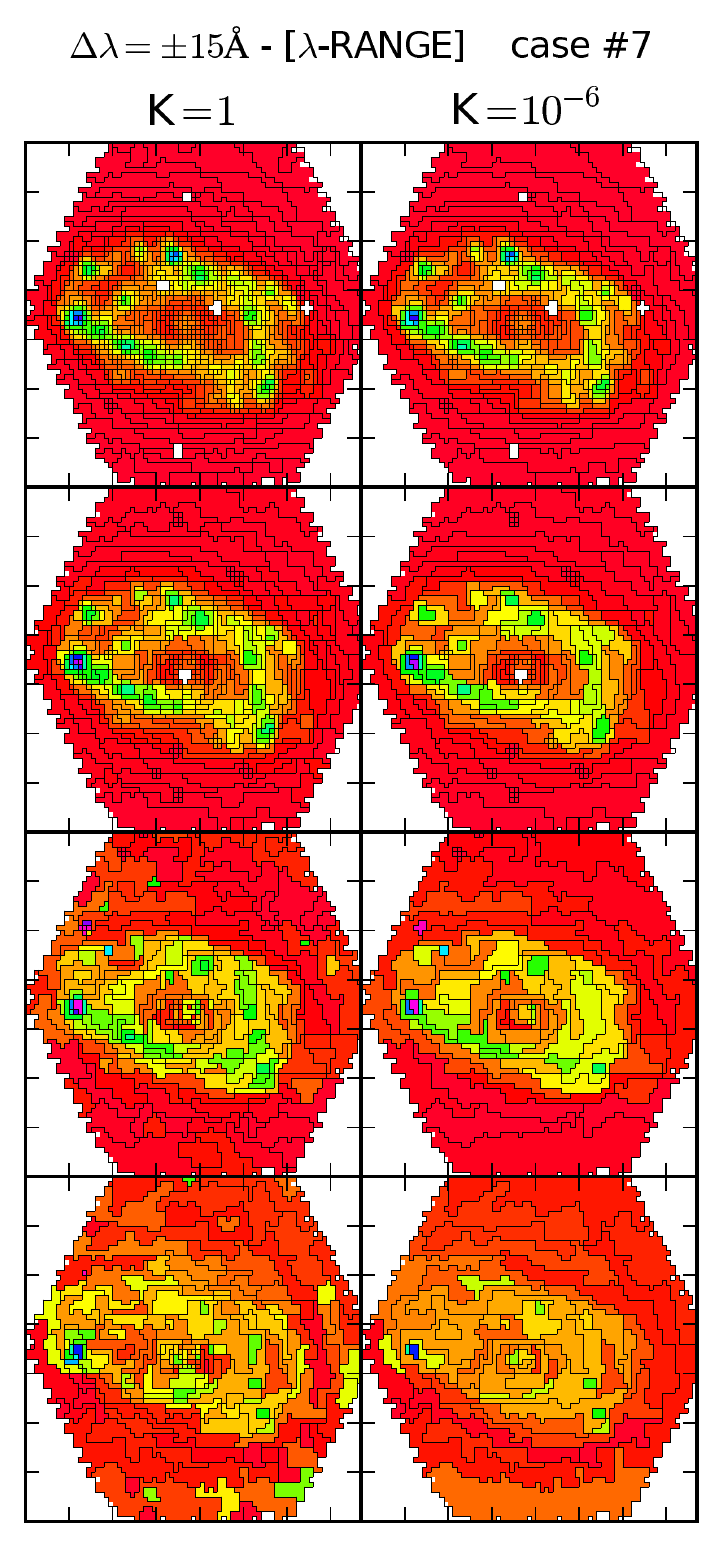}
~~ ~~ ~~ ~~
\caption
{
Results for our observational test cases based on CALIFA observations of \ngc. 
The structure of the plot is identical to that of Figure~\ref{fig_test_RGB} for the upper panels (test cases \#5 and \#6), which show the output of applying \batman\ to the intensity maps of the Balmer lines measured by \shifu\ on a spaxel-by-spaxel basis.
The lower panels show the results obtained for test cases \#7 and \#8, which are conceptually identical to \#5 and \#6, respectively, but are both multi-$\lambda$ tessellations of selected slices of the IFS datacube (see the description in Section~\ref{sec_obs} and Table~\ref{tab_tests}); it is impossible to represent the input signal in 2D for these cases.
The colour maps denote flux in standard CALIFA units, i.e. $10^{-16} (erg/s/cm^{2})$; they correspond to the \batman\ output in test cases \#5 and \#6 (upper panels), and to \shifu\ measurements posterior to the segmentation in test cases \#7 and \#8 (bottom panels).
}
\label{fig_test_galaxy}
\end{figure*}
%------------------------------------------------

\subsection{Large-scale morphology}
\label{sec_morphology}
% -------------------------------------------------------------------------

Figures~\ref{fig_test_RGB} and~\ref{fig_test_galaxy} show the results of applying \batman\ to our synthetic and astrophysical test cases, respectively, adopting the extreme values $K=1$ and $K=10^{-6}$ for the prior parameter, and considering both the full input data set as well as individual layers in order to compute the Bayesian evidence (see Table~\ref{tab_tests}).
From visual inspection, one can readily verify that, in almost all cases, the tessellations returned by the algorithm adapt extremely well to the structures that were present in the input data.

In the synthetic datacubes (Figure~\ref{fig_test_RGB}), \batman\ successfully recovers the different objects regardless of their topology.
The output tessellations present a variety of sizes and shapes that are driven by the input signal, and they show no preference for any given direction.
Comparing the results obtained for the `uniform' and `CCD-like' noise schemes, we conclude that the ability of the algorithm to trace the underlying structure is not very sensitive to the statistical characteristics of the noise. The minor differences that exist in the recovered tessellations can be explained in terms of the dissimilar signal-to-noise ratios.

When applied to the real data (Figure~\ref{fig_test_galaxy}), our algorithm manages to identify most of the clumps associated with physical {\sc Hii} regions, albeit with significant difficulty in the case of \hd.
In particular, among the \hd\ recovered maps, the poorest results are obtained for case \#5, where only the \hd\ intensity and its error map, as computed by \shifu, are used as input dataset.
Interestingly, the situation improves when \batman\ is run directly on the CALIFA COMBO dataset.
Even when only a narrow spectral region, `$\Delta=\pm 15$~\AA', around \hd\ is provided as input (case \#7), the ring-like structure where most of the emission comes from is prominently visible in the recovered signal.

By construction, \batman\ gathers together spaxels that contain the same information (compatible signal within the errors).
Hence, it is reassuring, but not surprising, that it successfully recovers the circle and the triangle in the R and G layers of the synthetic multi-images, as these regions do indeed feature exactly the same signal value.
However, this is not the case for the gradient in the B-band ellipse or the observations of \ngc, where the signal varies more or less smoothly across any layer.
Our segmentation model, described in Section~\ref{sec_idea}, does not consider the presence of gradients inside the regions. Therefore we expect the optimal tessellation to trace the isocontours perpendicular to the direction of any gradient that may exist in the signal. The size of the regions will be determined by the intensity of the gradient and the signal-to-noise ratio.
Roughly speaking, the size of the isocontours should be of the order of the typical uncertainties \ee\ within the region, so that the algorithm can tell that the signal is indeed different from the values in the immediate vicinity.

This expectation is consistent with the behaviour that can be observed in Figures~\ref{fig_test_RGB} and~\ref{fig_test_galaxy}.
The area of the synthetic multi-images covered by the ellipse is divided into concentric regions that roughly follow its shape. Most of the tessellations in \ngc\ tend to feature annuli that trace the isocontours of the considered signal.
It is worth noting that, in one of the astrophysical tests (\#5 and \#6), this refers to the measured intensities of the emission lines, whereas in the other (\#7 and \#8) it corresponds to the surface brightness at different wavelengths, including not only the emission lines but also the adjacent continuum.
In the central part of the galaxy, the signal-to-noise ratio is high.
There are many regions consisting of very few spaxels, and all the lines are clearly detected.
In the outskirts, where the signal is weaker, the regions are larger, and the overall normalization of the continuum plays a major role in setting their shape.

Since the morphology of the tessellation traces the spatial structure of the input data, the results depend on how these data are fed to the algorithm.
The first noticeable difference between the binning modes described in Section~\ref{sec_procedure} (and summarized in Table~\ref{tab_tests}) is that the `multi-$\lambda$' procedure yields a single tessellation for all layers. Conversely, the regions of each layer in the `monochromatic' mode have been obtained separately, and hence they are different and independent on one another.

In our synthetic problem, the circle, triangle, and ellipse are, by construction, truly independent objects, and they appear as such when \batman\ is run in the monochromatic mode.
In contrast, when the (R,G,B) layers are considered as a multi-image, the overlap regions are \emph{different} from the rest, as they would correspond to `yellow' and `purple' colours rather than pure `red', `green', or `blue'.
Hence, \batman\ separates these regions from the `main' objects (circle, triangle, and ellipse) according to the monochromatic definition.
Both solutions are equally correct, and which one is to be preferred depends on the specific scientific goals to be achieved.
It is entirely up to the user's judgement to decide what information should be considered relevant in order to decide whether two regions are `different'.
As an equivalent scientific example, the different layers in a multi-image may correspond to spatially-resolved maps of the \emph{mass} and/or the \emph{fraction} of young stars in a given galaxy, their age, metallicity, radial velocity, velocity dispersion, etc., and one can be interested in analysing all/some of these properties separately (`monochromatic' mode) or considering all of them at the same time (`multi-$\lambda$' mode).
The choice of the input data is a critical step of any analysis, and it will obviously play an important role in determining the solution found by the algorithm.

We also face such a decision in our specific examples of astrophysical test cases, where one may adopt different prescriptions with the ultimate goal of recovering intensity maps of the Balmer emission lines.
Is it more appropriate to make a first estimate on a spaxel-by-spaxel basis, and then bin the maps in order to increase the signal-to-noise ratio, or is it better to find an optimal tessellation based on the raw data and then carry out the measurements?
In either case, is it more convenient to consider each line separately, or all of them simultaneously?
Although it is certainly not our goal to answer these questions, it is interesting to compare several different approaches in order to illustrate the effect that this kind of choices has on the optimal tessellation that \batman\ returns.

When all the \shifu\ measurements are taken as input data (case \#6, Figure~\ref{fig_test_galaxy}), the output tessellation is largely driven by the \ha\ map, which has the highest signal-to-noise ratio.
Based on the information provided by this line alone, it is relatively easy to realize that two given spaxels are different if one of them belongs to an {\sc Hii} region and the other corresponds to diffuse emission, where the intensity is considerably lower.
The information contained in the weakest lines (\hg\ and \hd) is also used, but it carries a much lower weight, and it merely represents a minor-order correction.
If we run \batman\ separately on each map (case \#5, Figure~\ref{fig_test_galaxy}), the tessellation obtained for \ha\ is very similar to the multi-wavelength regime.
For \hb\ and \hg, the size of the regions increases due to the lower signal-to-noise ratio, but the overall structure is still the same.
For \hd, the algorithm adapts to the structures that are present in the input data, but these are severely affected by the noise.
This is clearly a disadvantage in our test problem, since we know that the signals in the different layers are not independent.
However, this may not be true in the general case (e.g. if there was a signal that was present \emph{only} in \hd), and then it could well be possible that it is more adequate to consider the noisy dimensions separately in order to maximise their weight on the tessellation.

When applied directly to the IFS data (case \#7 and \#8, Figure~\ref{fig_test_galaxy}), the algorithm is able to recover the ring-like structure in all cases, even when only a 30-\AA\ interval around \hd\ is taken as input.
The central regions of the galaxy are always divided into relatively small regions, whereas the outer parts are tessellated in roughly concentric rings.
The overall morphology is actually fairly robust, and the main differences between the different prescriptions are related to the region sizes.
In fact, the tessellations obtained from each of the individual intervals (case \#7), as well as from the full data set (case \#8), are qualitatively similar, at variance with the results based on the synthetic test problems or the \shifu\ maps.

\subsection{Features on small scales}
% -------------------------------------------------------------------------

\subsubsection{Random fluctuations}
\label{sec_fluctuations}

In our synthetic tests (see Figure~\ref{fig_test_RGB}), it is evident that \batman\ recovers many more aggregations of one (or a few) spaxels when only a single `layer' is analysed (`monochromatic' mode) than when all layers in the multi-image are considered simultaneously (`multi-$\lambda$').
These regions are statistically significant according to \batman's criterion, i.e. the posterior probability ratio~\eqref{eq_posterior_candidate} indicates that it is more likely that the enclosed signal is statistically different from that of the adjacent regions.
We do know, however, that these regions may be `real' only in the sense that the input data are indeed different as the result of statistical fluctuations of the noise\footnote{This is the so-called \emph{look-elsewhere} effect: a large fluctuation at a random place is extremely unlikely, but a few large fluctuations \emph{somewhere} in the sample are indeed expected.}.
By adding Gaussian noise, some individual spaxels may deviate significantly (e.g. more than three sigma) from the surrounding neighbours, even if the underlying signal is constant.
It is in fact relatively likely that two or more (up to a few) adjacent spaxels happen to deviate in the same direction by a significant amount (e.g. $-2 \sigma$ and $-2.5 \sigma$), and, depending on the adopted prior, BaTMAn may classify them as an independent region, statistically different from the surrounding ($\pm 1 \sigma$) `background'.

As we increase the number of dimensions, the significance of each individual layer decreases.
For example, a fluctuation $\Delta = (3.5\sigma_1, -0.5\sigma_2, 1.1\sigma_3, -1.2\sigma_4, 0.2\sigma_5, 0.5\sigma_6)$ in six dimensions yields $\chi^2 = \sum_i \frac{\Delta_i^2}{\sigma_i^2} = 3.5^2 + (-0.5)^2 + 1.1^2+ (-1.2)^2 + 0.2^2 + 0.5^2 = 15.44$, and it has considerably less statistical significance than the $3.5$-$\sigma$ fluctuation in the first layer alone.
In terms of $\chi^2$, it is completely equivalent to a $1.6$-$\sigma$ fluctuation in all layers ($6\times1.6^2 =15.36$).
Moreover, if there is indeed a signal over a small region, all spaxels are biased in the same direction, whereas this is very unlikely for a random fluctuation: more precisely, the probability that all $n_\lambda$ layers in \nreg\ spaxels fluctuate in the same (e.g. \emph{positive} or \emph{negative}) direction by sheer chance decreases as $2^{-\nreg n_\lambda}$, and random aggregations of a few adjacent spaxels are thus much harder to form in the `multi-$\lambda$' case.

For analogous reasons, a true signal is also more difficult to detect: a clear difference in one layer becomes less and less significant if it is diluted among a plethora of other channels carrying unrelated information (and/or noise).
In reality, it is far from trivial to discriminate weak signals from random fluctuations, and this is again a challenge left for the final user.
In our tests based on astronomical data, the situation is further complicated by the large differences in S/N from one line to another as well as the potential presence of observational artefacts.
For the \ha\ maps, based on either \shifu\ measurements or CALIFA data, including information from the other Balmer lines tends to reduce the number of regions in the areas where the signal is stronger (i.e. within the ring-like structure), preferentially removing regions with one or two spaxels.
In the outer parts, though, a number of single-spaxel regions arise in the multi-wavelength tessellations due to the presence of outliers in \emph{any} of the other layers.
Although a careful study would be required in order to assess whether these outliers are physical, it is our preliminary impression that most of the single-spaxel regions in the galaxy outskirts are due to artificial features in the data.
For all the Balmer lines other than \ha, the monochromatic tessellations always yield much larger regions than the multi-wavelength result, because of their much smaller signal-to-noise ratio.

\subsubsection{Final number of regions}

Regardless of their ultimate physical nature (random fluctuations or real signal), the small-scale structure of the output tessellation is determined by the number, size, and arrangement of the smallest regions that are identified by the algorithm as statistically significant.

As explained in Section~\ref{sec_code}, the iterative merging procedure continues as long as there is any candidate tessellation $R_{c,n-1}$ where the posterior probability ratio~\eqref{eq_posterior_candidate} is larger than unity, or, equivalently,
\begin{equation}
\prod_{\lambda=1}^{n_\lambda}
(H_\lambda-L_\lambda)
\frac{ e^{-\frac{ (\mu_{A\lambda}-\mu_{B\lambda})^2 }{ 2(\sigma_{A\lambda}^2+\sigma_{B\lambda}^2) }} }
     { \sqrt{2\pi(\sigma_{A\lambda}^2+\sigma_{B\lambda}^2)} }
> K 
\end{equation}
Thus, the combined prior parameter $K$ regulates the contrast that is required in order to consider that two regions correspond to different signal values.
The effect of considering different values of the combined prior parameter $K$ is illustrated in Figures~\ref{fig_test_RGB} and~\ref{fig_test_galaxy} by depicting the results obtained for $K=1$ and $K=10^{-6}$ in all our example test cases.
A decreasing value of $K$ essentially increases the probability of merging regions, and some of the small fluctuations discussed above are removed as a result of a less strict compatibility criterion.

In terms of the prior probability distributions~\eqref{eq_prior_theta} and~\eqref{eq_prior_R}, regulated by the parameters $k_\lambda$ and $k_{\rm R}$, respectively, a low value of $K$ is closer to the `non-informative' (infinite) improper uniform prior, and therefore it means that we have little previous knowledge about the signal and/or that we prefer a segmentation with as few regions as possible.
Both statements are absolutely equivalent, both formally as well as in practice.
$K \sim 1$ uses the information contained in $H_\lambda$ and $L_\lambda$ to minimise the probability that two regions with physically different signal are merged together.
Lower values of this parameter reduce the contamination associated to random fluctuations in the input data and increase the signal-to-noise ratio at the expense of missing the weakest genuine signals.
The optimal value of $K$ is problem-dependent, and it can only be decided upon trial and error.
Fortunately, our results suggest that there are relatively little differences between adopting values as extreme as $K=1$ and $K=10^{-6}$, both in synthetic as well as in real data.

To sum up, we can conclude that \batman\ is able to adapt to the structures present in the data, both at large and small scales.
The resulting tessellation depends slightly on the adopted priors and, most importantly, on the precise choice of the input data set.
Therefore, it is important to devote some time to investigate these issues as a preliminary step of any scientific analysis.
The local signal-to-noise ratio and the precise value of the combined prior parameter $K$ (the only free parameter of the algorithm) affect the number and size of the regions identified by the algorithm, especially when gradients are present, but neither of them has a significant effect on the overall morphology of the tessellation.
Other aspects, such as the actual shape of the underlying structure, or the statistical properties of the noise, do not seem to play a major role.

%------------------------------------------------
\begin{figure*}
\centering
\includegraphics[height=.4\textheight]{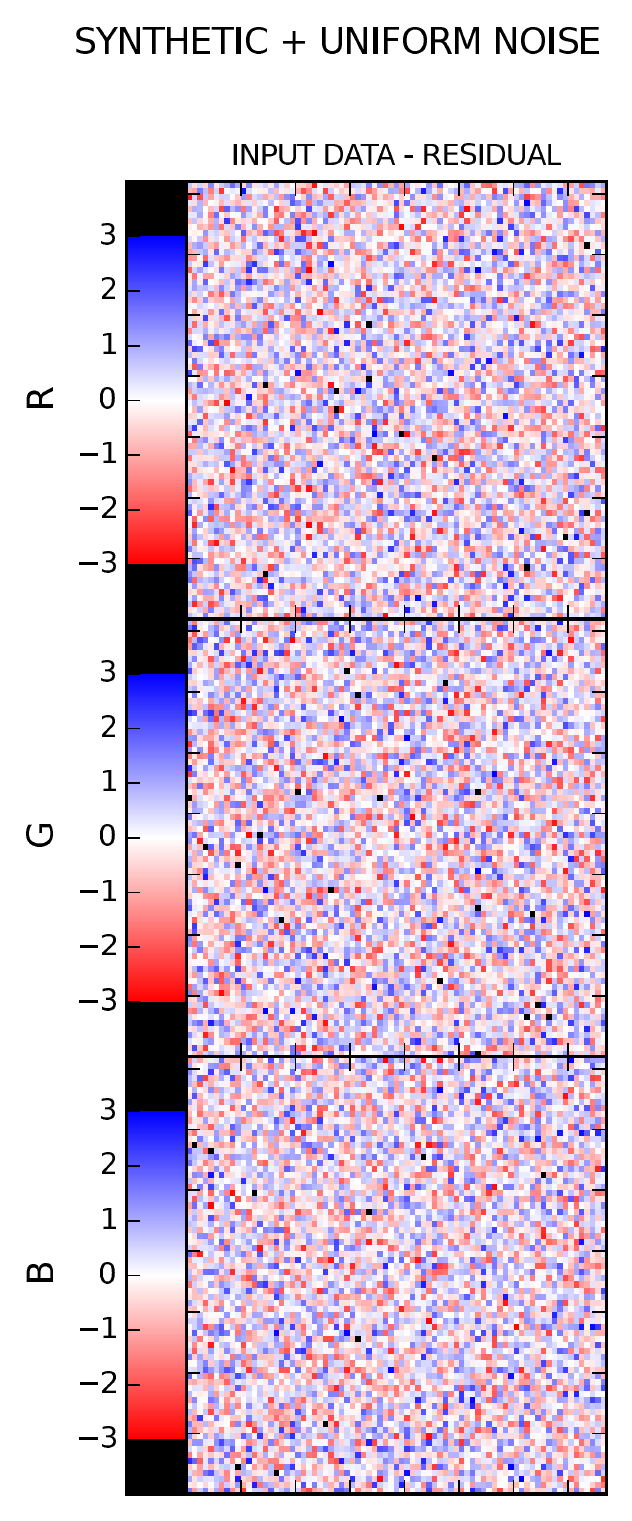} ~~ ~~ ~~
\includegraphics[height=.4\textheight]{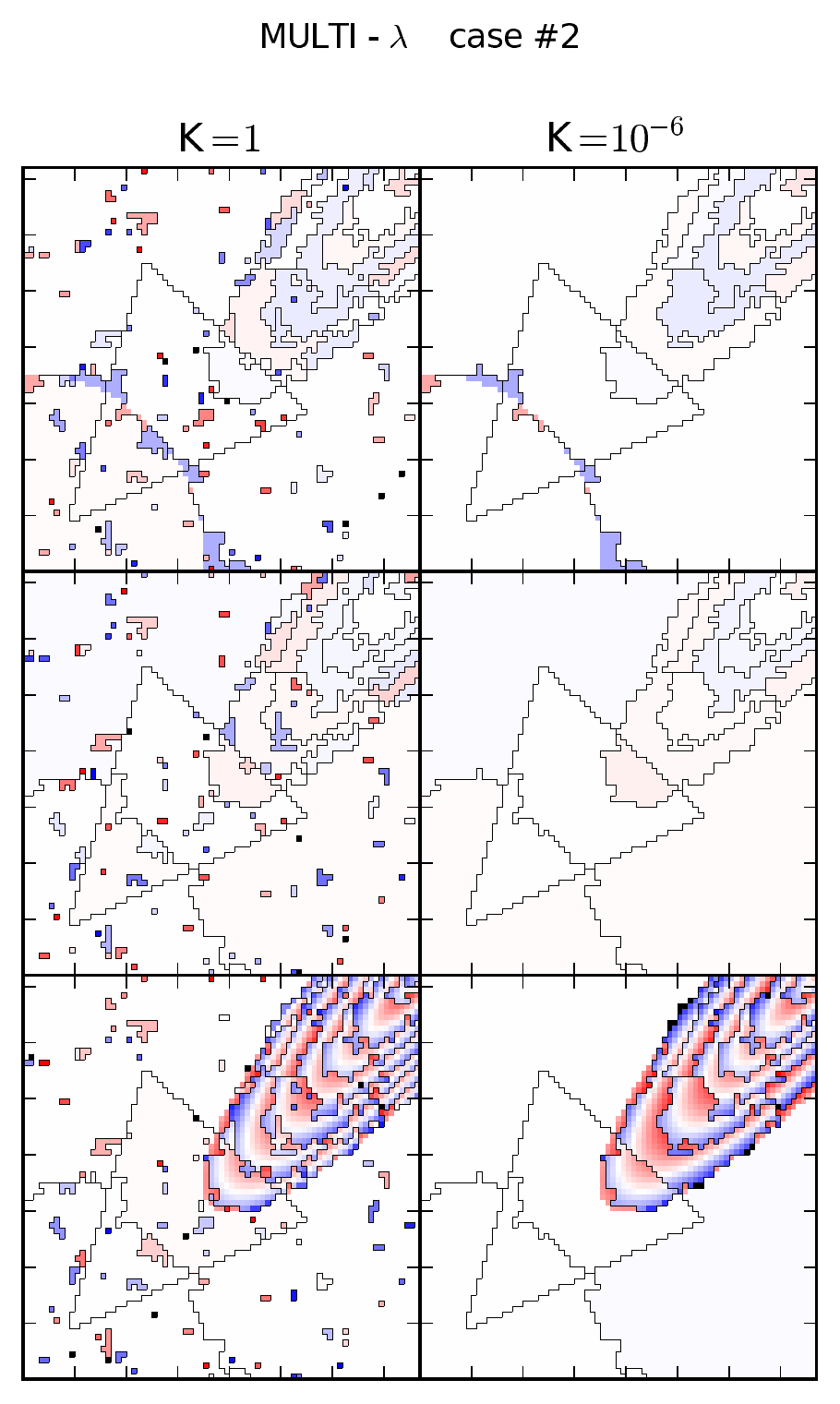}~~
\includegraphics[height=.4\textheight]{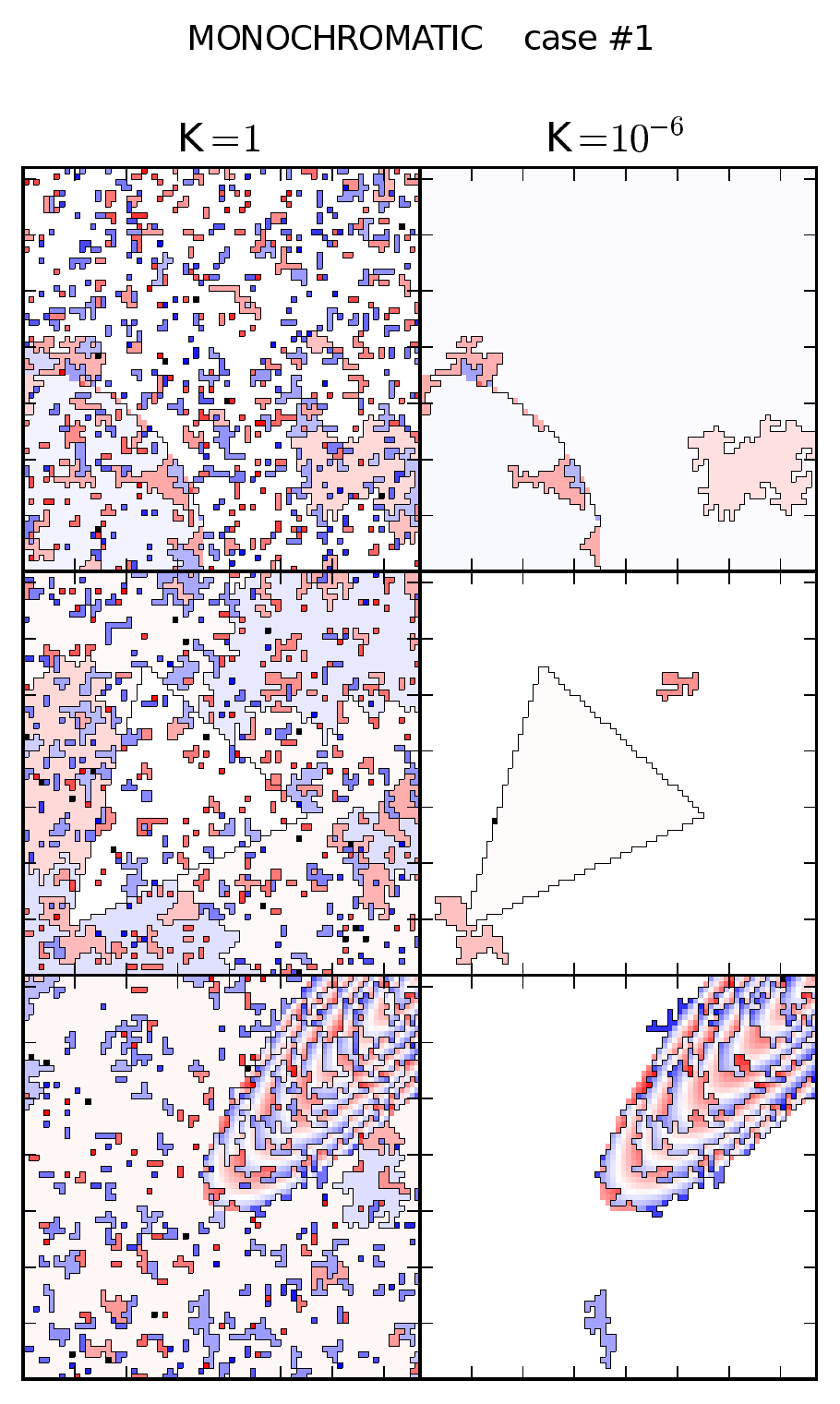}\\[3mm]
\includegraphics[height=.4\textheight]{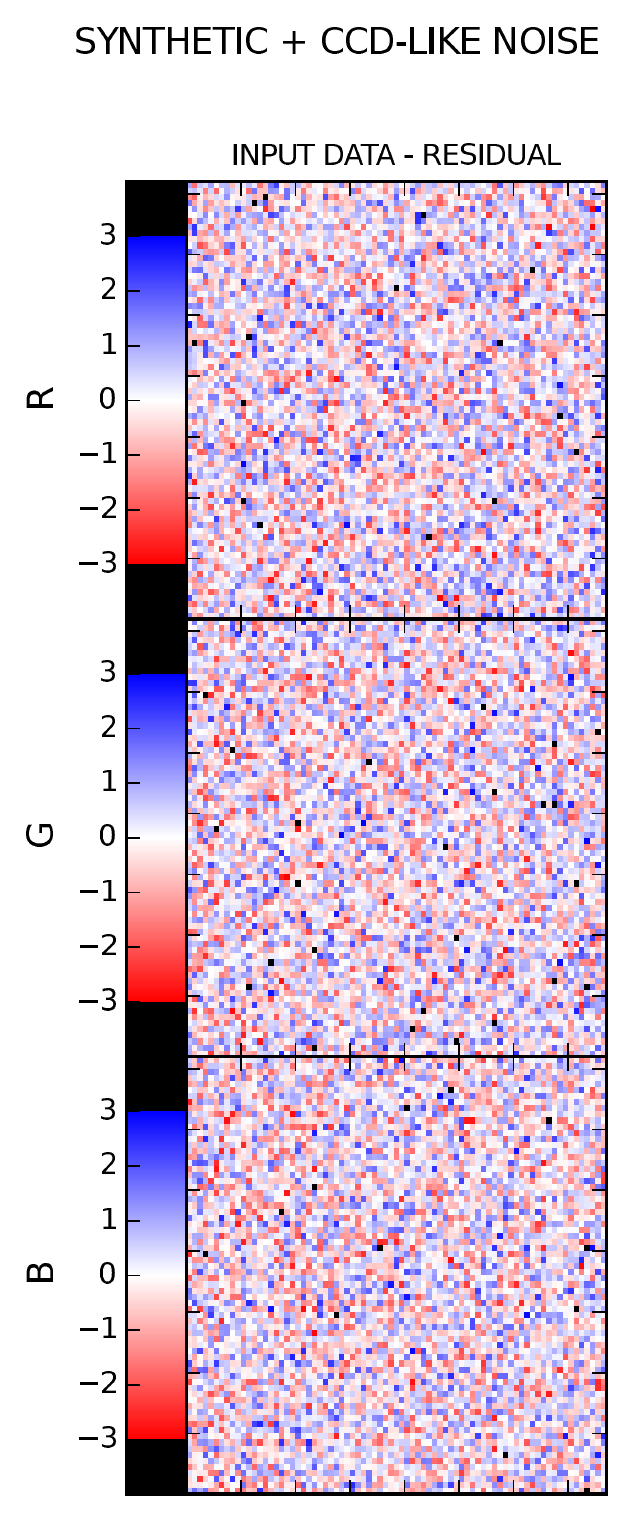} ~~ ~~ ~~
\includegraphics[height=.4\textheight]{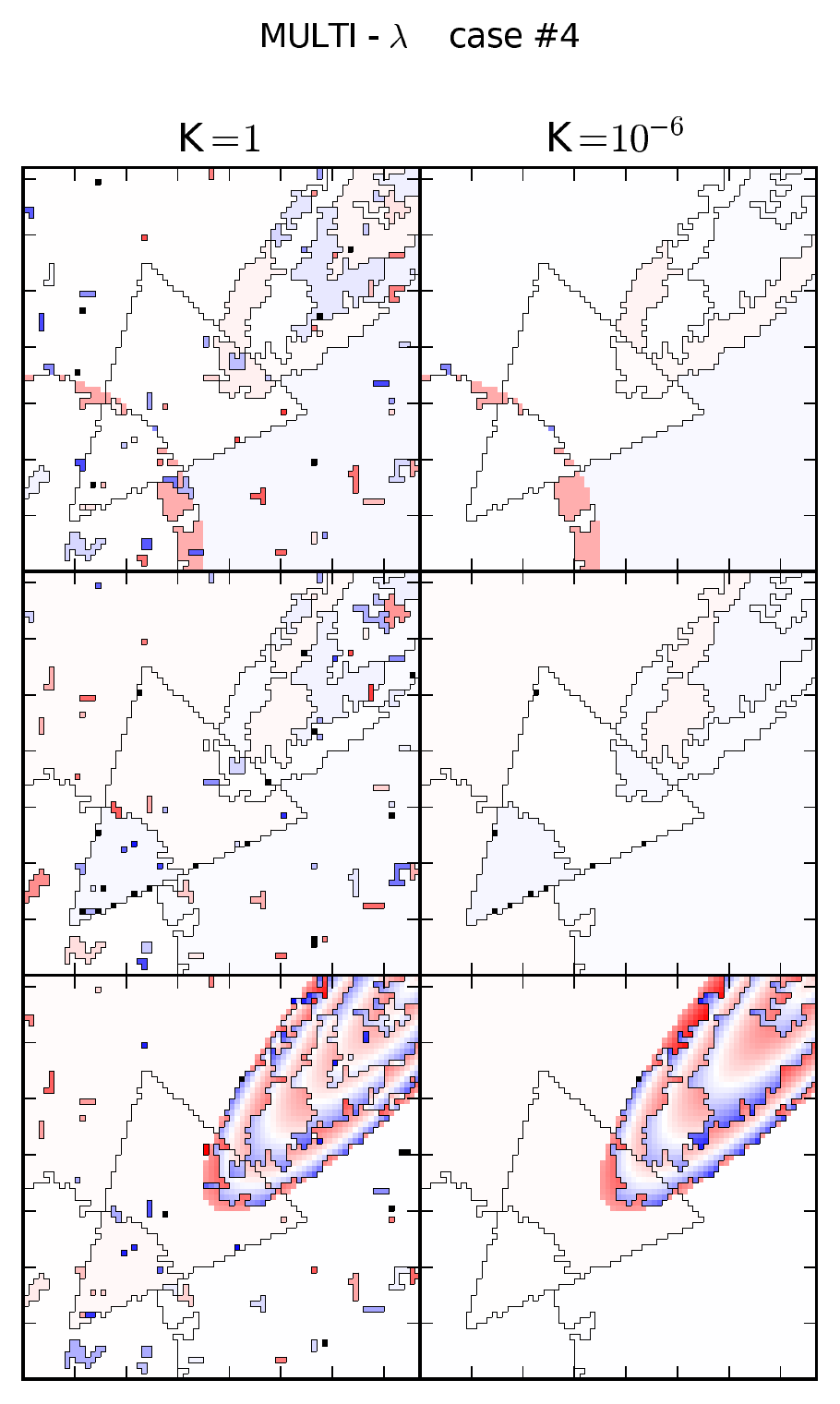}~~
\includegraphics[height=.4\textheight]{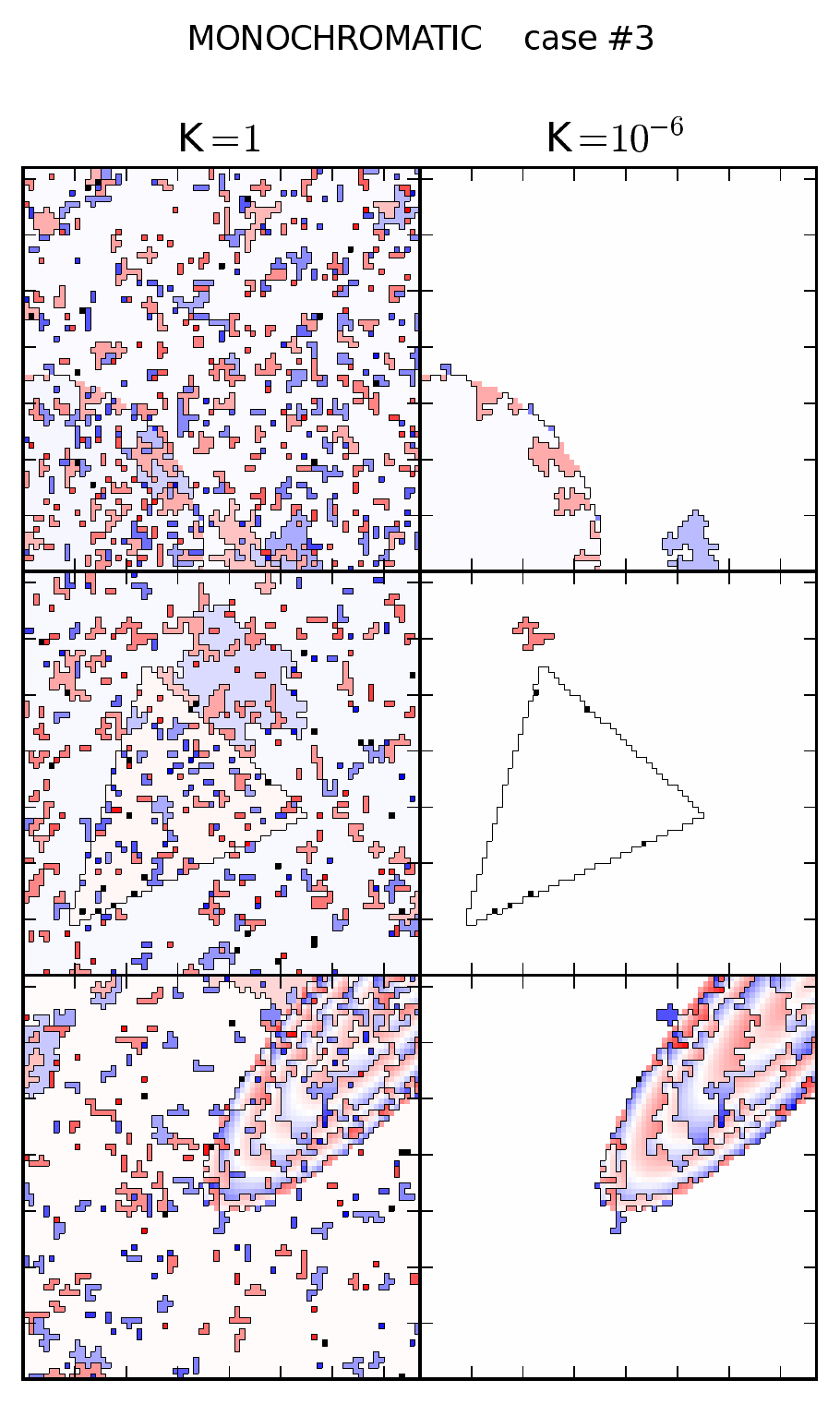}\\[3mm]
\caption
{
Maps of the residuals $\residual = \frac{ \mu_{r_{ij}\lambda} - \signal }{ \ee }$ for our synthetic test cases.
The structure of the plot is identical to Figure~\ref{fig_test_RGB}.
}
\label{fig_error}
\end{figure*}
%------------------------------------------------
\begin{figure*}
\centering
\includegraphics[height=.35\textheight]{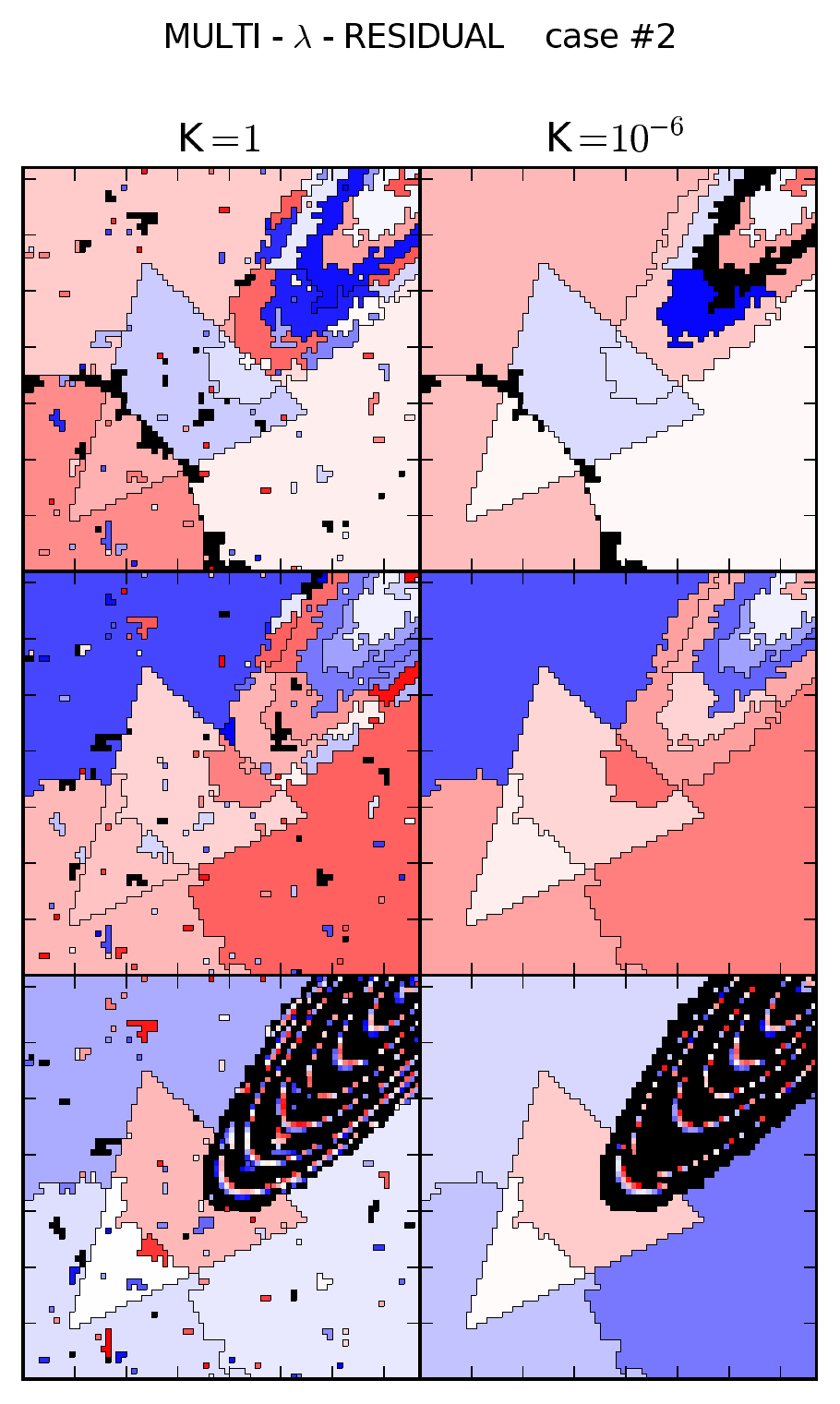}
\includegraphics[height=.35\textheight]{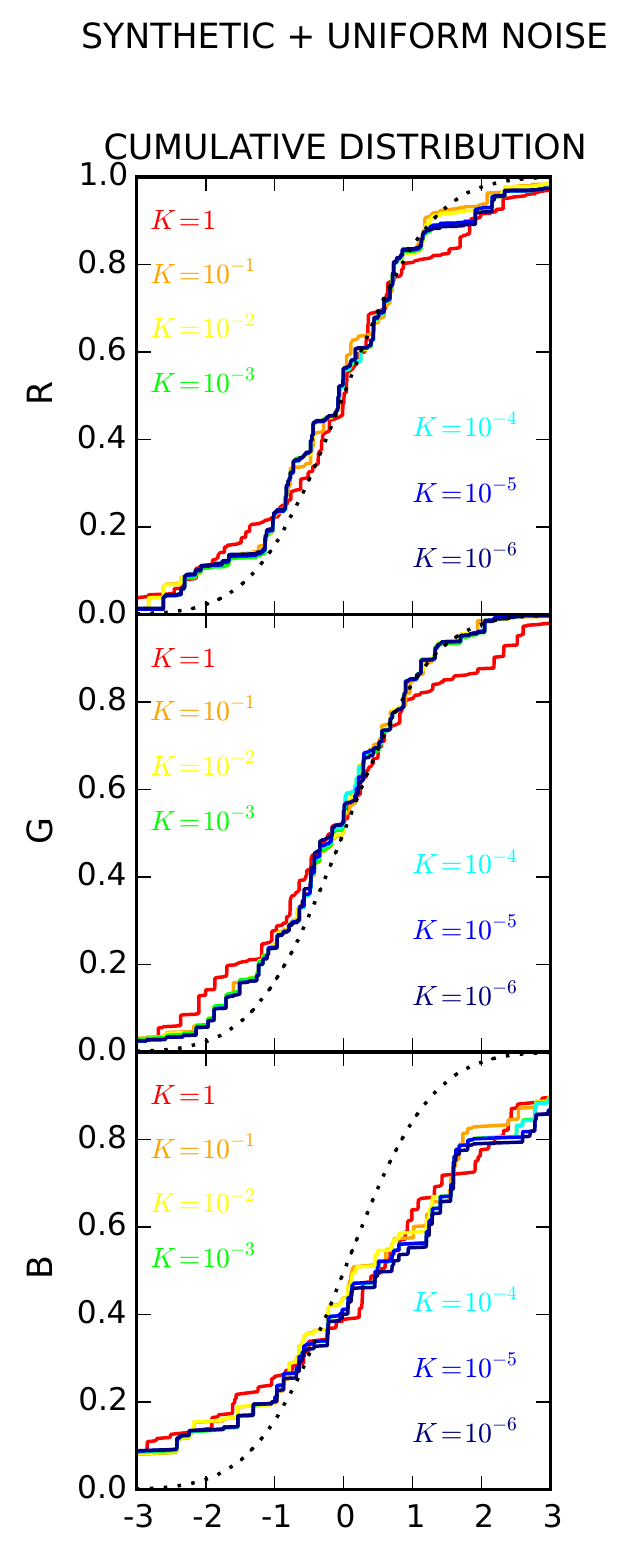}
\includegraphics[height=.35\textheight]{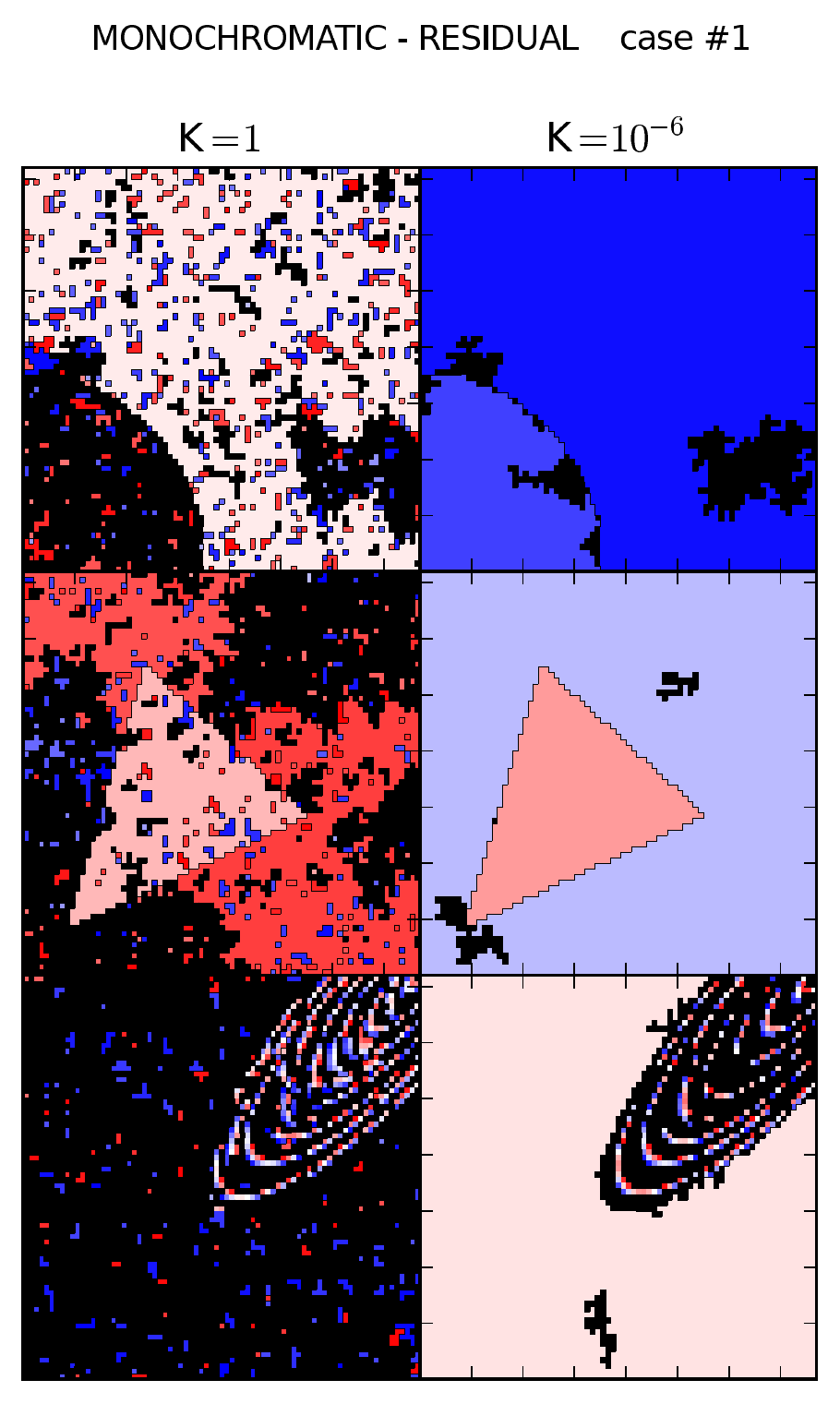}
\includegraphics[height=.35\textheight]{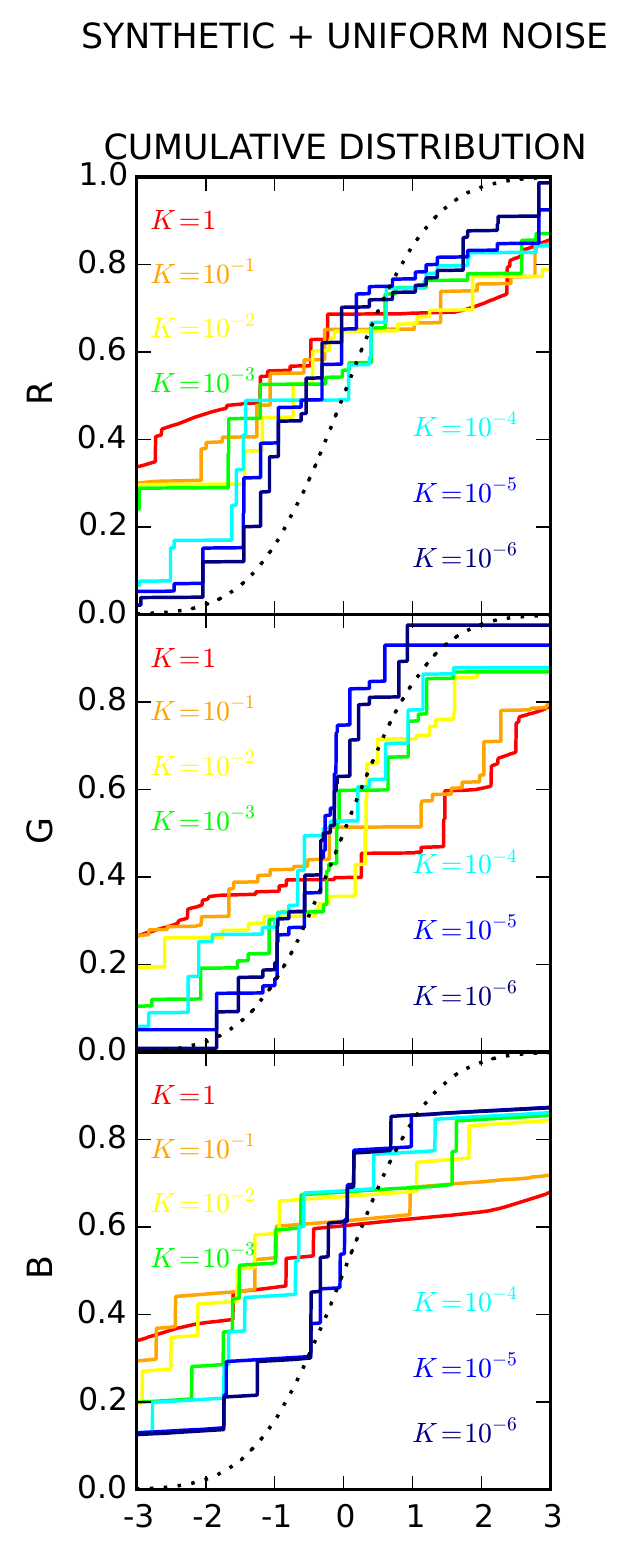}\\[3mm]
\includegraphics[height=.35\textheight]{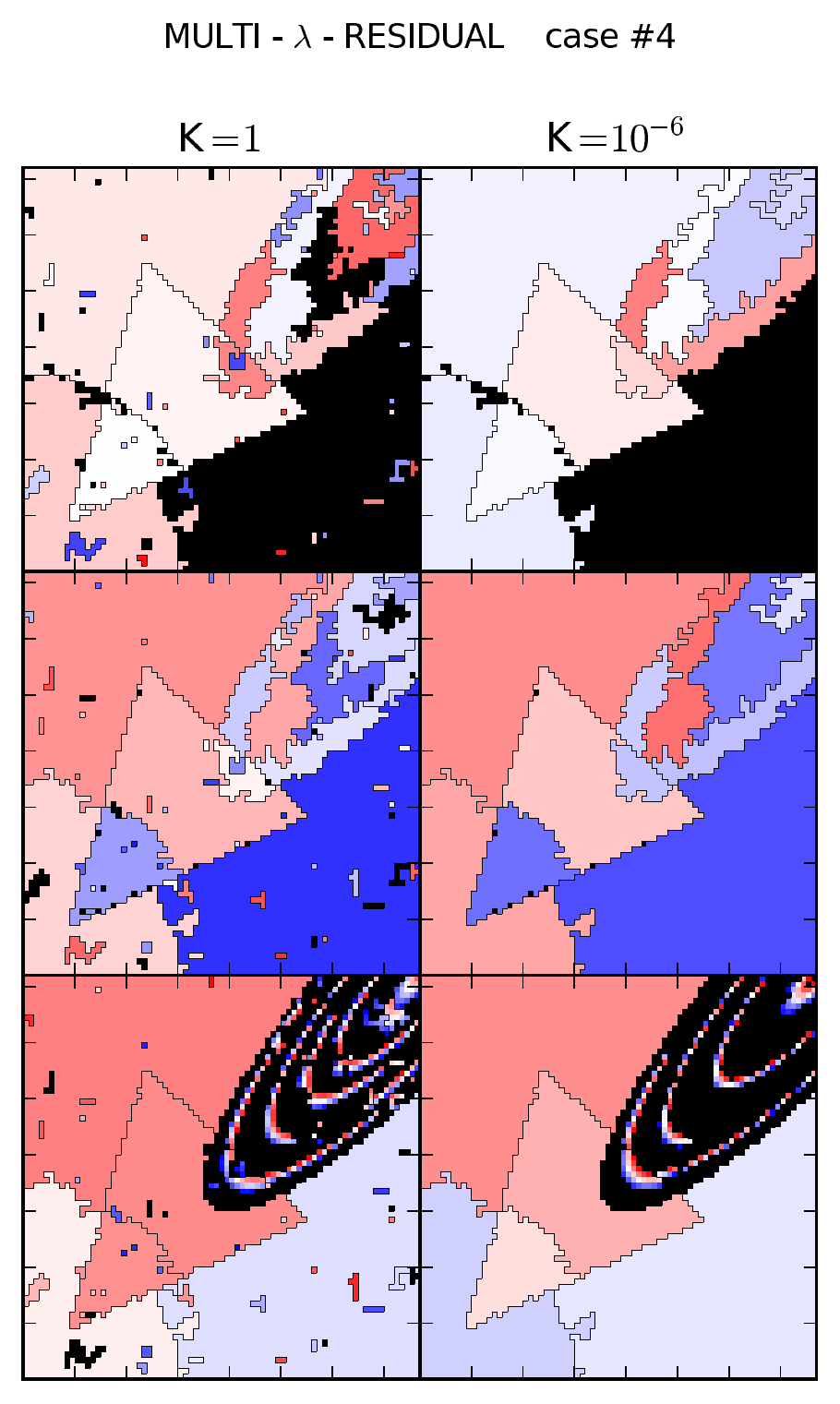}
\includegraphics[height=.35\textheight]{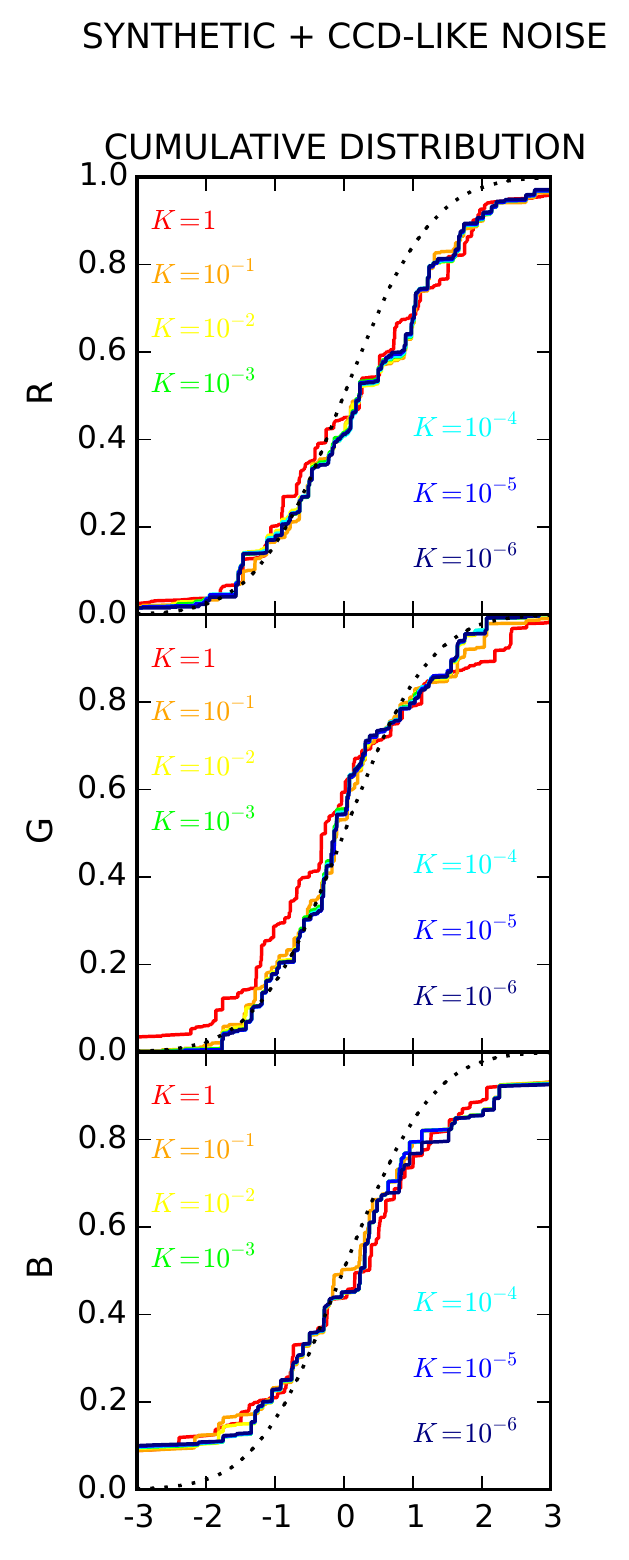}
\includegraphics[height=.35\textheight]{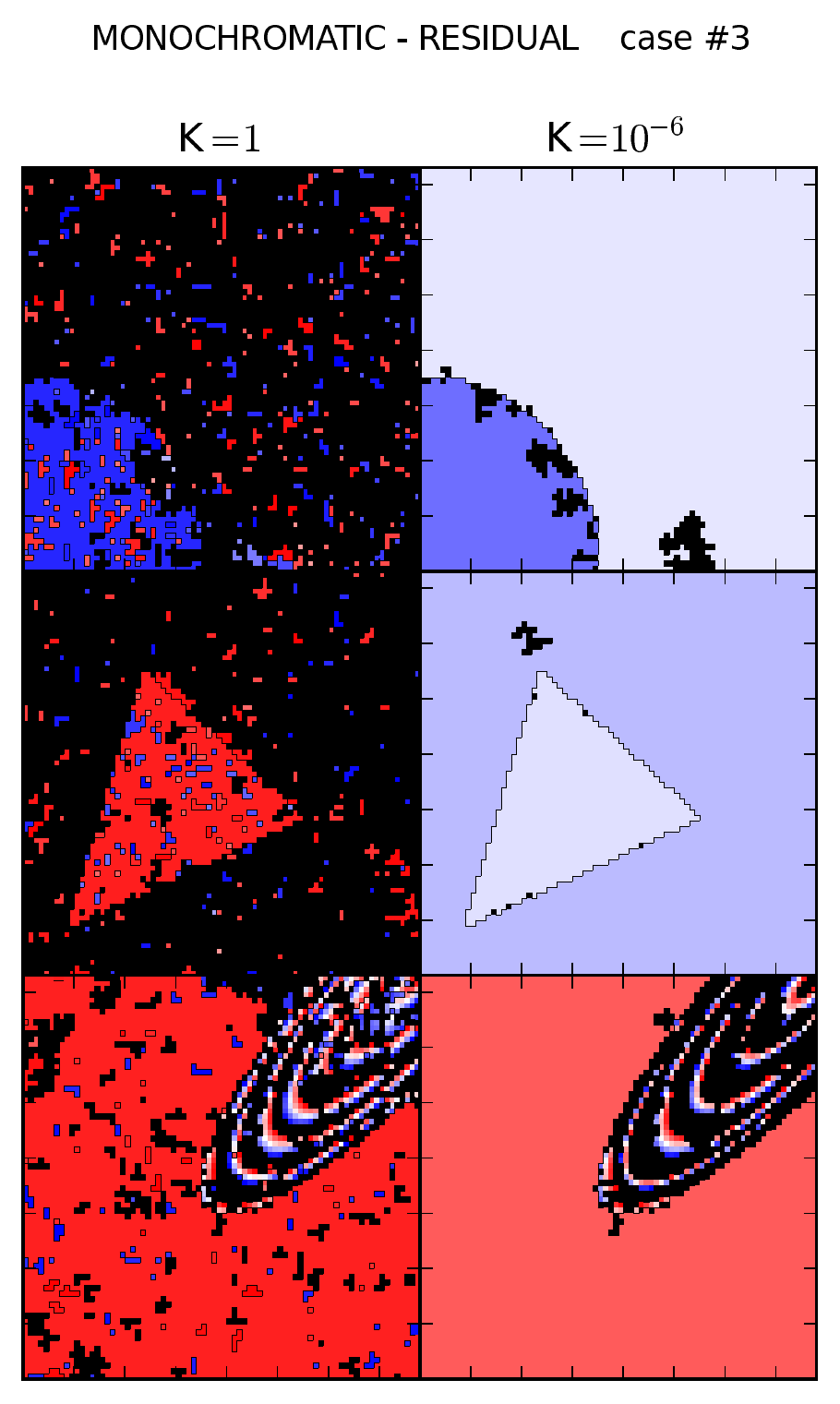}
\includegraphics[height=.35\textheight]{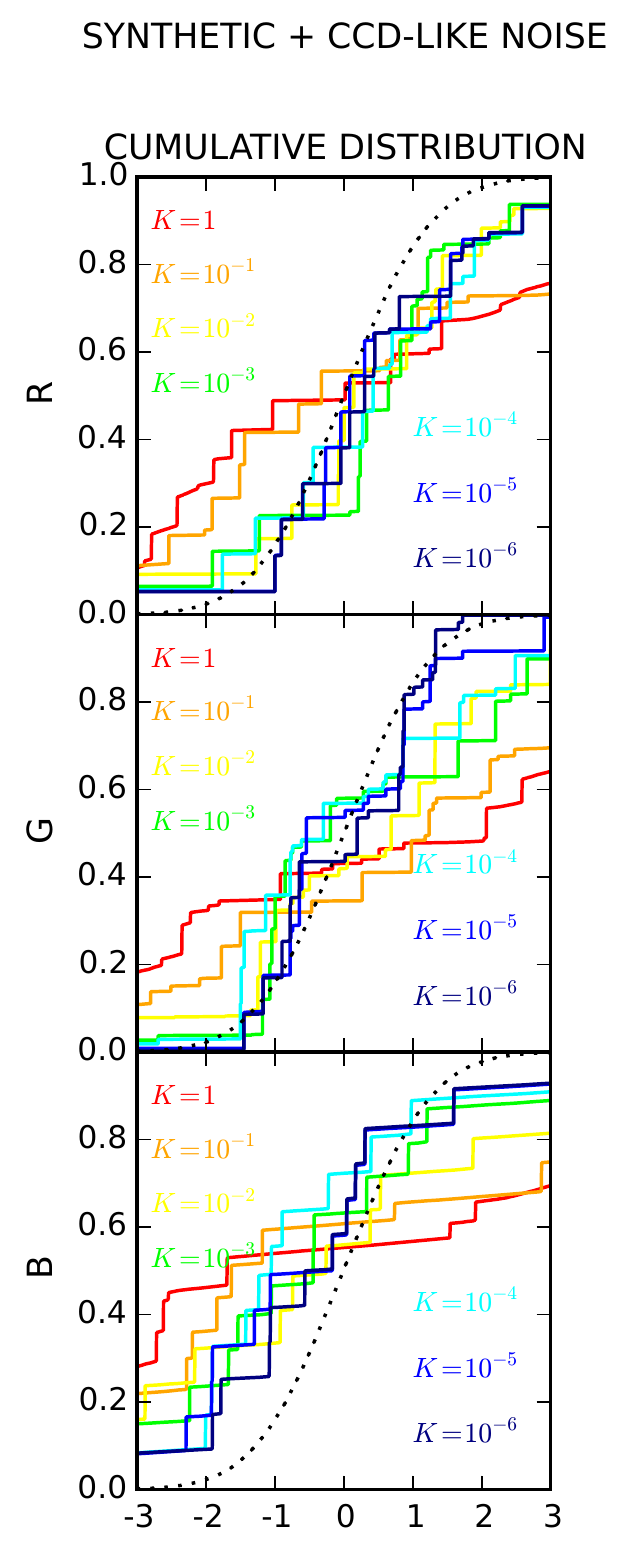}
\caption
{
Maps and cumulative distributions of the estimated residuals $ \residualB = \frac{ \mu_{r_{ij}\lambda} - \signal }{ \sigma_{r_{ij}\lambda} } $ for our synthetic test cases.
The structure of the plot is similar to that of Figure~\ref{fig_test_RGB} except for the missing input data, for which $\residualB = \residual$ (left panels of Figure~\ref{fig_error}).
We include the cumulative distributions of \residualB\ to study the behaviour of the formal errors \sg\ returned by the algorithm, where different colours correspond to different values of the prior parameter $K$ (see labels in the figure).
In order to improve the statistics, the data plotted in these cumulative curves correspond to the results of applying \batman\ to 10 random realizations of the noise.
Black dotted curve corresponds to the unbinned input data for the same 10 realizations, where we correctly recover the Gaussian distribution of the noise.
Ideally, all the other curves should follow as close as possible this distribution; a more extended range of values of \residualB\ indicates that the formal errors \sg\ underestimate the true uncertainty (i.e. the rms difference between $\mu_{r_{ij}\lambda}$ and \signal).
}
\label{fig_sigma}
\end{figure*}
%------------------------------------------------
\subsection{Quality of the reconstruction}
\label{sec_reconstruction}
% -------------------------------------------------------------------------

One of the reasons to tessellate a multi-image is to obtain a better reconstruction of the underlying signal, especially in the low signal-to-noise regions.
In order to calibrate the improvement with respect to the original input data, we study in Figure~\ref{fig_error} the distribution of the residuals for our two synthetic test cases, where \signal\ is known, in terms of the input errors \ee,
\begin{equation}
 \residual = \frac{ \mu_{r_{ij}\lambda} - \signal }{ \ee }
 \label{eq_residual1}
\end{equation}
with $\mu_{r_{ij}\lambda} = \mu_{r\lambda}$ for all spaxels $\{i,j\}\in r$, i.e. $r_{ij} = r$.
For the input multi-image, where $\mu_{r_{ij}\lambda}=\xx$, the residuals \residual\ follow by construction independent Gaussian distributions with zero mean and unit dispersion.
Therefore, the residual maps on the left column of Figure~\ref{fig_error} are statistical realizations of white noise for all layers, without any spatial structure, both for the `uniform' and `ccd-like' schemes.

As can be seen in the other maps, the posterior mean values \mm\ obtained for every \batman\ region are representative of the signal values within the region.
In the R and G layers, where the signal within the circle, the triangle, and the background is indeed constant, \mm\ is very close to the true value ($\residual\simeq 0$, white colour in the residual maps) as long as the region is correctly identified.

In some cases, though, a small number of spaxels may be assigned to a region where the underlying signal is actually different: in our synthetic tests, this only happens near the boundaries of the R circle, which are difficult to trace due to the low signal-to-noise ratio ($S/N=1$).
One may expect that such misclassifications yield $\residual\sim\mathcal{O}(1)$.
If the contrast between any two areas is larger than \ee\ (as it occurs e.g. in the triangle in the G band), the difference between adjacent spaxels on each side of the boundary is statistically significant. It is thus very unlikely that they become connected by our iterative procedure before merging with the other surrounding spaxels that belong to the correct region.
Although these errors are indeed possible, they are restricted to a few isolated spaxels in the G layer of the `ccd-like' case, where $S/N=3.5$ (see Table~\ref{tab_sn}).

As discussed above, \batman\ can also identify regions around sufficiently high (a few sigma) random fluctuations on small scales when high values of the combined prior parameter $K$ are adopted, especially in the `monochromatic' case.
Again, random fluctuations larger than $\residual\sim\mathcal{O}(1)$ are statistically very unlikely, and therefore the residuals in the affected spaxels are of this order (red/blue colours in Figure~\ref{fig_error}).

Errors of similar magnitude also occur in the presence of gradients, where continuous variations of the signal \signal\ are physically present in the data, whereas \mm\ is assumed, by definition, to be constant within every region.
Even if the isocontours are correctly traced by the tessellation, and the value of \mm\ is representative of the average signal within the region, we expect, following the same argument applied to the misclassifications, that $\residual\sim\mathcal{O}(1)$. Otherwise, the deviant spaxel would be associated to the adjacent bin of correspondingly higher or lower intensity.
The effect of gradients is illustrated by the ellipse in the B band, where a characteristic red-white-blue pattern is clearly visible within every \batman\ region.

We thus conclude from Figure~\ref{fig_error} that the expected values of the posterior probability \mm\ returned by our algorithm provide a more accurate representation of the underlying signal than the original input data over most of the multi-image.
Only for the `catastrophic' failures discussed above, residuals of the order of $\residual\sim\mathcal{O}(1)$ may be found.

Another important aspect is the quality of the formal errors \sg\ derived from equation~\eqref{eq_variance} in Section~\ref{sec_param}.
In the ideal case, where all the underlying assumptions are met, the estimated residual
\begin{equation}
 \residualB = \frac{ \mu_{r_{ij}\lambda} - \signal }{ \sigma_{r_{ij}\lambda} }
 \label{eq_residual2}
\end{equation}
should follow a normal distribution.
However, the presence of gradients in the signal and the identification of spurious regions will cause that the formal errors \sg\ reported by \batman\ should be treated with caution and merely taken as a (realistic) lower limit to the actual uncertainty in the recovered signal.

%------------------------------------------------
\begin{figure*}
\centering
\includegraphics[height=.33\textheight]{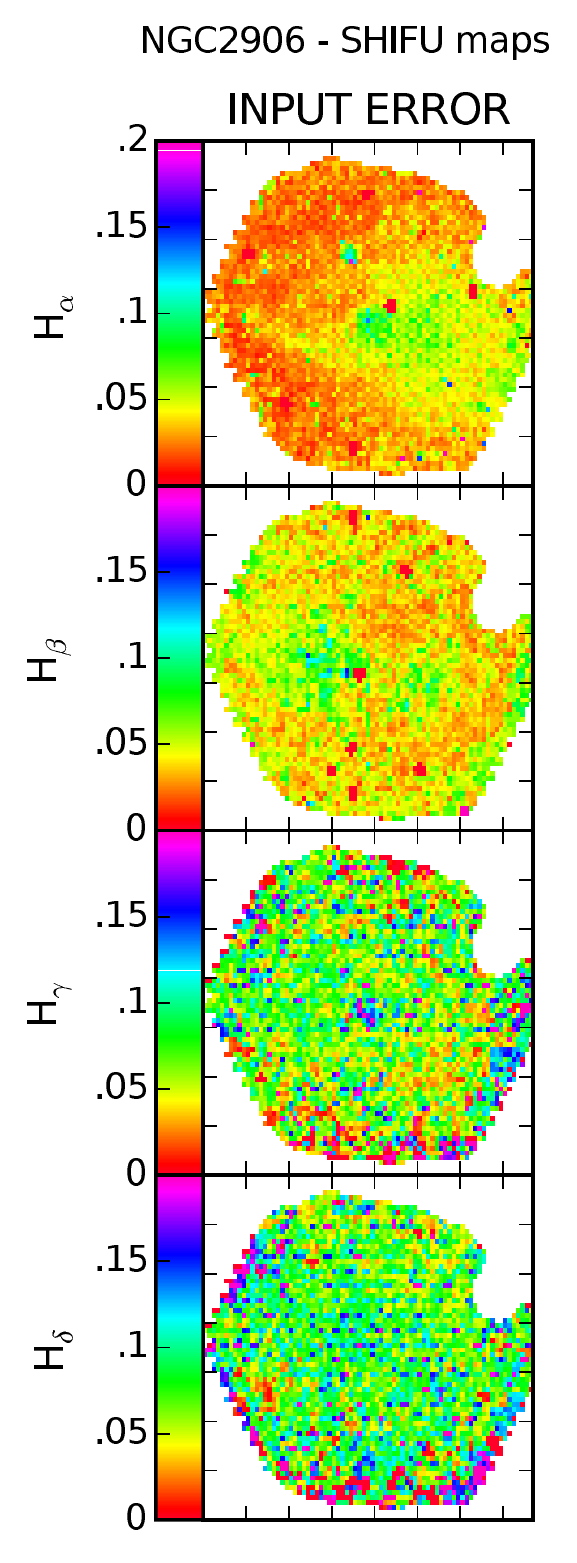}~~ ~~
% \hfill
\includegraphics[height=.33\textheight]{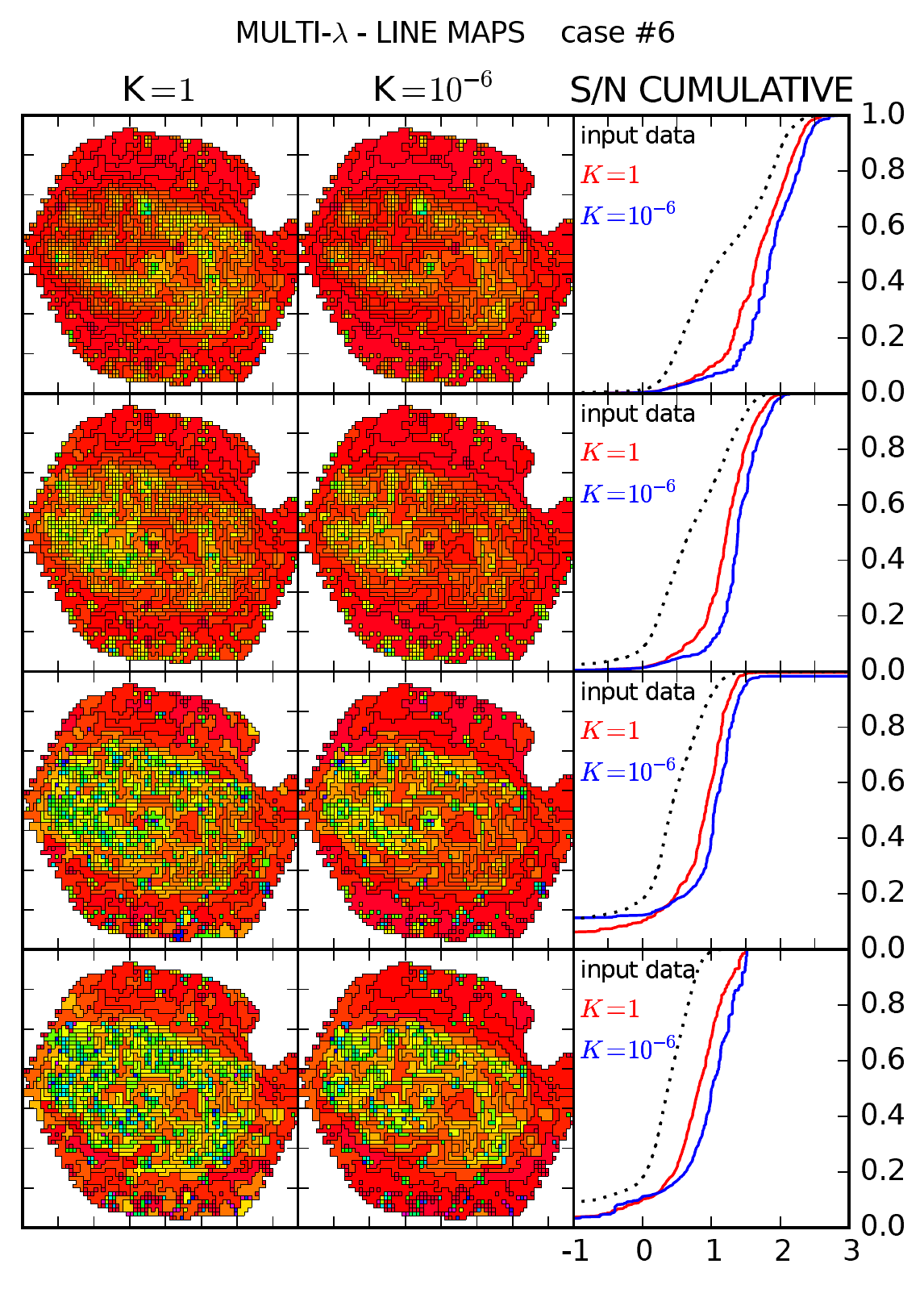}~~
\includegraphics[height=.33\textheight]{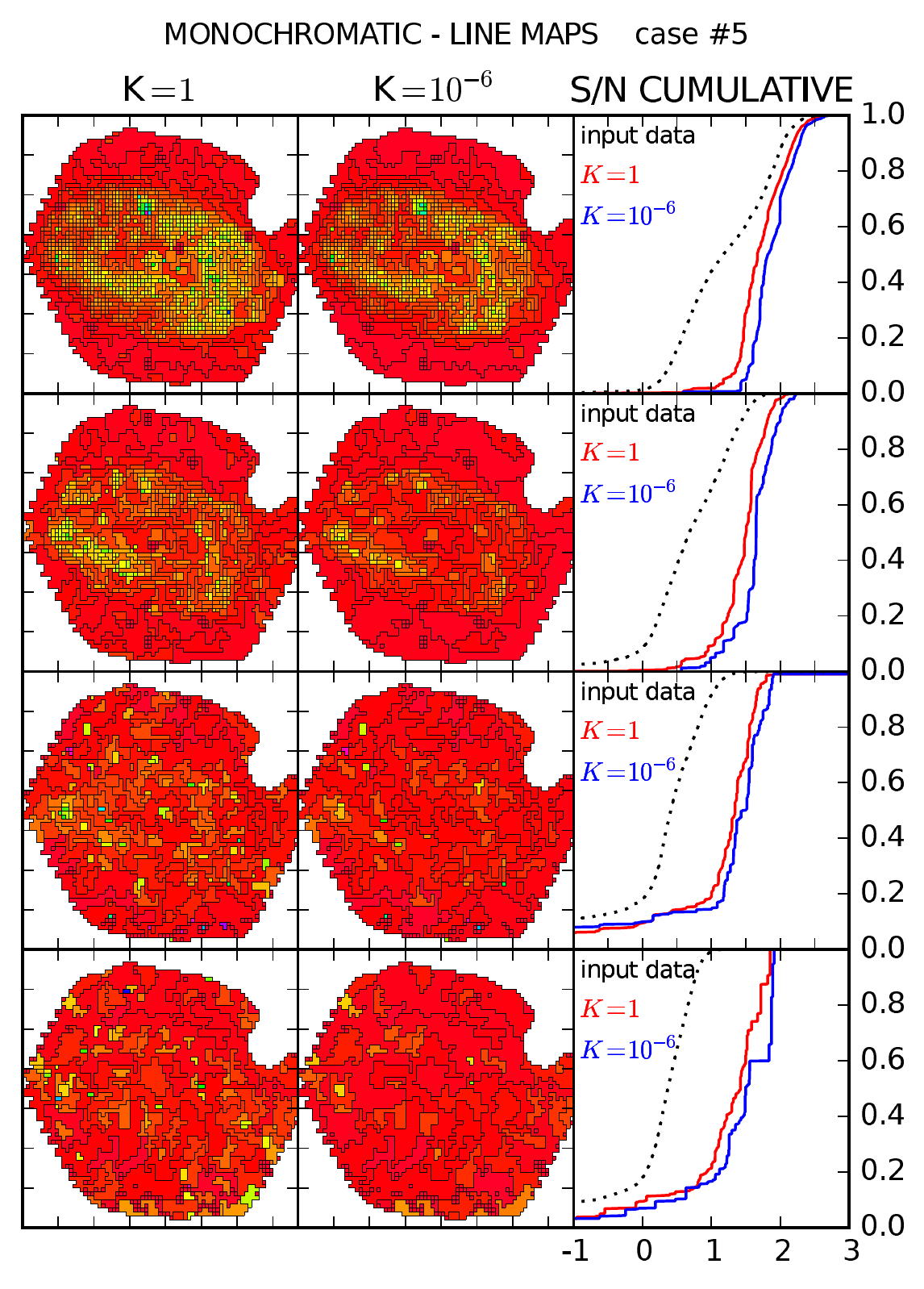}\\[3mm]
% \hfill
~~ ~~ ~~ ~~ ~~ ~~ 
\includegraphics[height=.33\textheight]{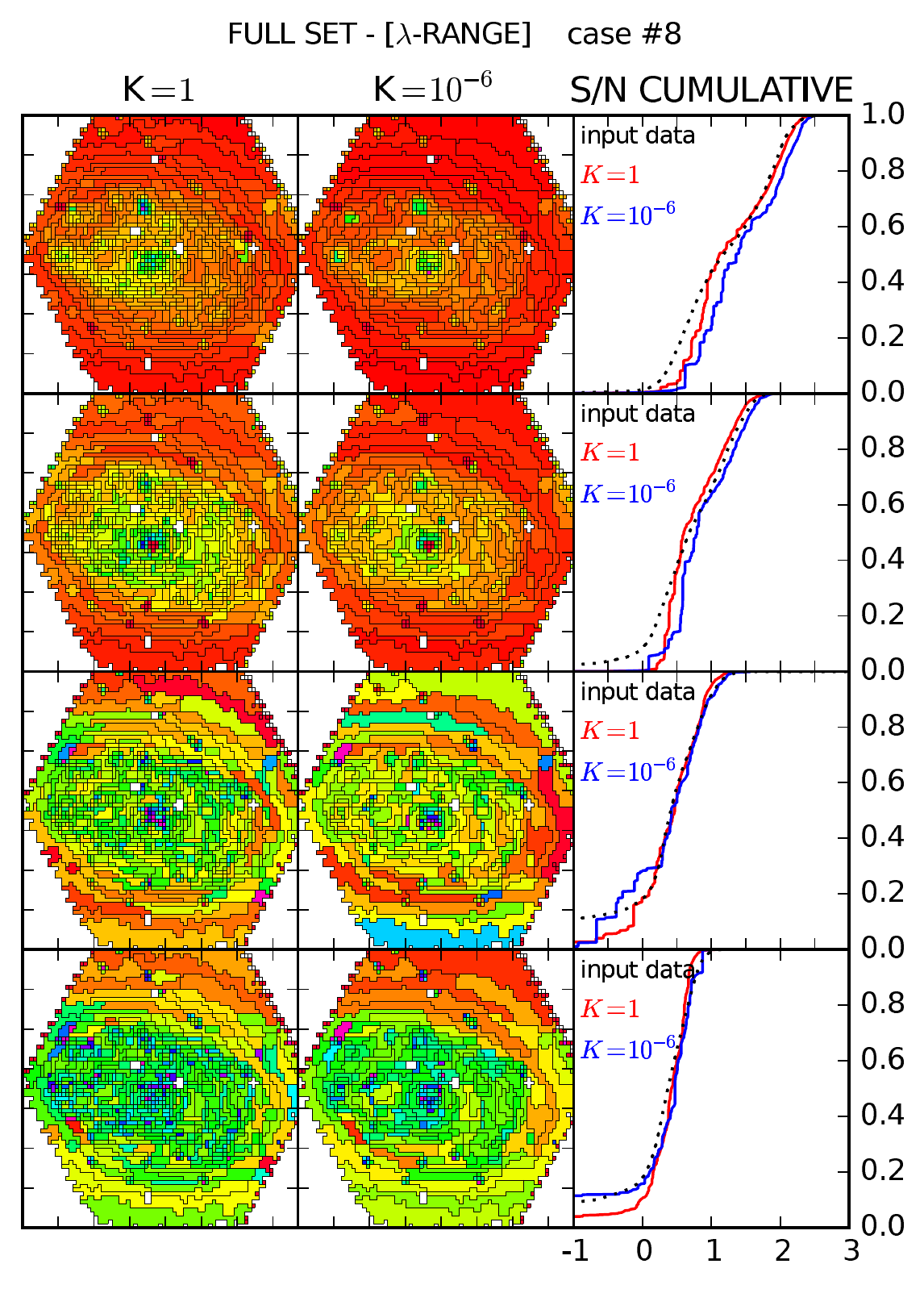} ~~
\includegraphics[height=.33\textheight]{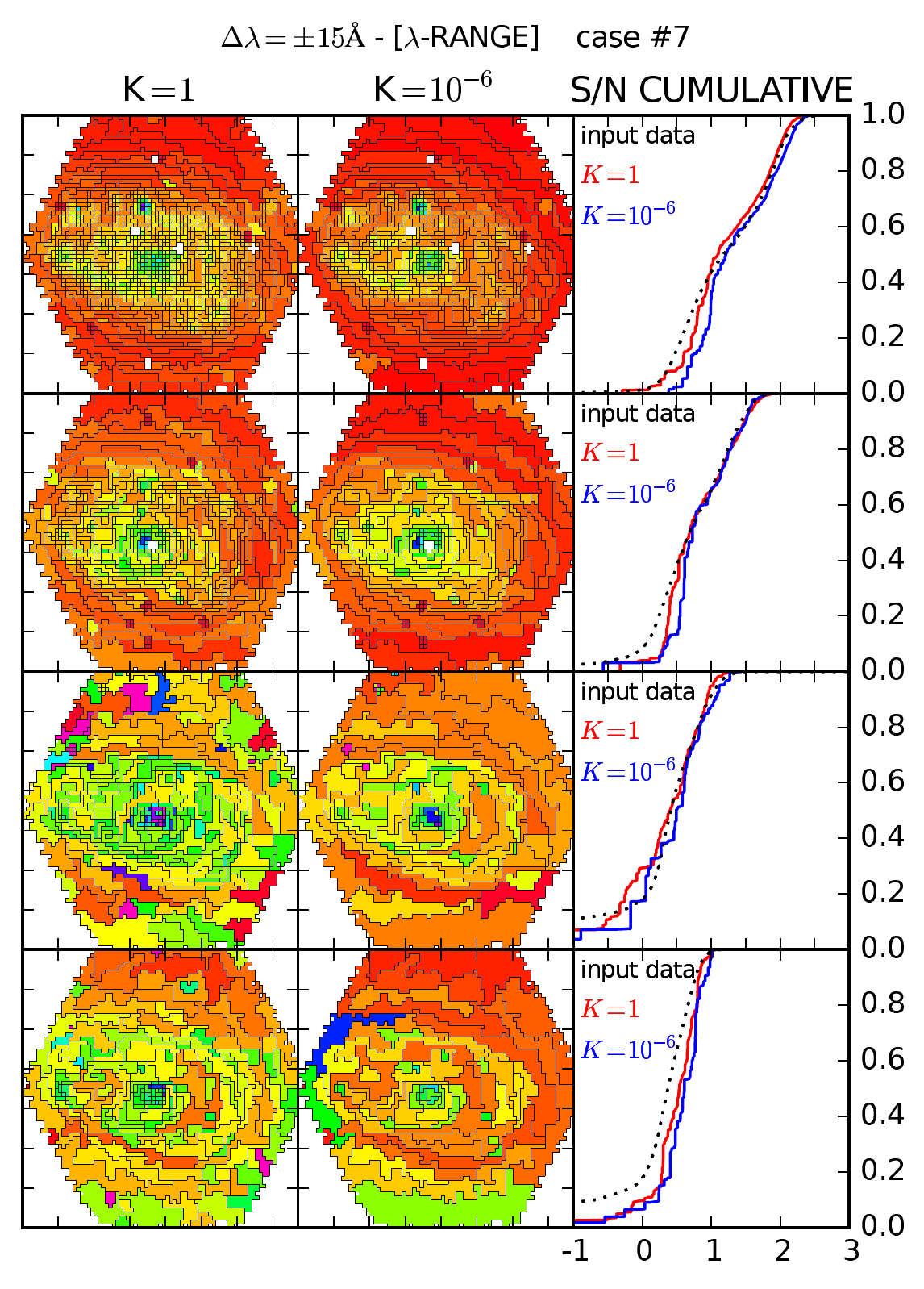} ~~
\caption
{
Results of applying \batman\ to our observational test cases, error maps and cumulative distributions of the `recovered' signal-to-noise ratios. The structure of the plot is identical to Figure~\ref{fig_test_galaxy} except for the inclusion of the cumulative distributions. Red and blue cumulative curves correspond to $K=1$ and $K=10^{-6}$ results. 
The dotted black curve corresponds to the \shifu\ maps used as input for test cases \#5 and \#6 and ploted in the left column.
The colour maps correspond to flux errors for every line in standard CALIFA units, i.e. $10^{-16} (erg/s/cm^{2})$. The signal-to-noise ratio is represented in logarithmic scale.
}
\label{fig_error_S/N}
\end{figure*}
%------------------------------------------------

A more quantitative assessment is provided in Figure~\ref{fig_sigma}, where we plot colour maps of the estimated residual \residualB. We also present histograms of its cumulative distribution (based on 10 independent realizations of the random noise) for different values of the combined prior parameter $K$.
Using our synthetic tests we find that, for $K<10^{-3}$ (lines of green-blue colour), the statistical distribution of the estimated residuals \residualB\ is \emph{similar, but not exactly equal} to the expected normal distribution.
For $K=1$, values of $|\residualB| > 3$ (black areas in the colour maps) are common in the `monochromatic' case and relatively frequent in the `multi-$\lambda$' analysis.
In the latter (and/or for $K<10^{-3}$), they represent a small fraction of the multi-image and therefore they have a barely noticeable effect on the cumulative distribution.

The most important departure with respect to the ideal behaviour is due to the presence of gradients in the data.
Although the algorithm manages to identify regions where the underlying signal is \emph{approximately} constant, equation~\eqref{eq_variance} is based on the assumption of strict equality, and physical variations within the region (of the order of \ee) are not taken into account in the error budget.
As can be seen in Figure~\ref{fig_sigma}, this source of systematic error may result in \sg\ underestimating the true residuals in any given spaxel by as much as a factor of the order of two, which arguably constitutes a reasonable upper limit, albeit not necessarily a worst-case scenario.

In order to investigate a more realistic example, let us now try to address the quality of the reconstruction in our astrophysical test cases.
As in any practical application, the underlying signal is unknown, and the true solution is not available.
In the top panels of Figure~\ref{fig_error_S/N} we compare the error maps \ee\ of the \shifu\ spaxel-by-spaxel measurements with the formal errors \sg\ recovered by \batman, both in the `monochromatic' (case \#5) and the `multi-$\lambda$' (case \#6) binning modes.
For test cases \#7 and \#8, we plot on the bottom panels of Figure~\ref{fig_error_S/N} the errors $e_{r\lambda}$ reported by \shifu\ when the regions defined by \batman\ from the CALIFA data are given as input.

In all cases, we do observe that the formal errors decrease (i.e. the colour maps become redder) and their distribution becomes more uniform thanks to the segmentation.
Although the improvement is certainly not surprising in the top panels (\sg\ decreases roughly as the square root of the number of spaxels in region $r$), it is an expected but non-trivial result in the bottom panels.
In fact, the errors in the line measurements performed by \shifu\ after the \batman\ segmentation ($e_{r\lambda}$) are much more similar to the results of the spaxel-by-spaxel analysis (\ee) than suggested by equation~\eqref{eq_variance}.

This is due to a combination of different non-linear effects.
The most important is arguably the correlation between the noise in adjacent spaxels in CALIFA data, for which \batman\ assumption of statistical independence is a poor approximation.
The propagation of the errors through the \shifu\ pipeline takes it into account through a correction factor that attains values of the order of $\sim 2-3$ for regions with $\sim 8-66$ spaxels \cite[see Appendix A in][for details]{Garcia-Benito+15}.
Other issues, such as potential misclassifications, gradients, and the heterogeneous quality of the observational data (including the error estimates) across different spaxels also contribute to increase the uncertainty.

Yet, the bottom panels in Figure~\ref{fig_error_S/N} indicate a significant reduction of the errors $e_{r\lambda}$ in the low signal-to-noise areas (e.g. the boundaries) of the multi-image, where large (often annular) regions are identified by \batman.
This improvement is consistent with the intensity maps displayed in Figure~\ref{fig_test_galaxy}, where the level of random fluctuations in these regions is substantially reduced compared to the spaxel-by-spaxel results shown in the top left panels.
In areas where the S/N of the CALIFA datacubes is high, our algorithm tends to return regions consisting of one or a few spaxels, and the results (both the intensity and its error) are virtually unchanged with respect to the spaxel-by-spaxel measurements.

We also investigate in Figure~\ref{fig_error_S/N} the changes in the cumulative distribution of the \emph{estimated} signal-to-noise, comparing $\xx/\ee$ (spaxel-by-spaxel measurements) with $\mm/\sg$ in the \batman\ output (cases \#5 and \#6) and with $x_{r\lambda}/e_{r\lambda}$ (\shifu\ measurements on \batman\ regions; cases \#7 and \#8).

First of all, we would like to argue that the interpretation of these quantities is not straightforward, and some of the first impressions conveyed by the histograms in Figure~\ref{fig_error_S/N} are grossly misleading.
In particular, they might seem to suggest at first sight that the best strategy to tackle this particular problem is to run \batman\ on the \shifu\ measurements, as this approach leads to the strongest reduction in the formal errors \sg\ and the largest increase in the estimated S/N (especially for \hd\ in the `monochromatic' case).

This is certainly at odds with our discussion in Section~\ref{sec_morphology}, based upon visual inspection of the maps in Figure~\ref{fig_test_galaxy}.
The original S/N in this particular case is so poor that \batman\ is barely able to even recover the large-scale morphology of the underlying signal.
As a consequence, the regions of the tessellation contain spaxels where the signal may vary considerably, and the formal errors are thus expected to be underestimated by about a factor of two, according to our previous test (Figure~\ref{fig_sigma}).
As mentioned above, the correlation between the noise of adjacent spaxels represents a further source of uncertainty, leading to an additional underestimation of the error that grows logarithmically with the number of spaxels in the region under consideration \citep{Garcia-Benito+15}.

The improvement of $x_{r\lambda}/e_{r\lambda}$ (cases \#7 and \#8) with respect to the spaxel-by-spaxel measurements $\xx/\ee$ looks much more modest, but we think it is more realistic.

When the signal-to-noise ratio is relatively high (e.g. $S/N>5-10$, inner parts of the galaxy), the CALIFA data are good enough to distinguish fairly minor differences between adjacent spaxels.
Since \batman's goal is to keep all the information that is present in the multi-image, the resulting regions tend to be small.
Due to the reduced number of spaxels and the presence of correlated noise, the measurement errors $e_{r\lambda}$ associated to these regions are only slightly smaller than \ee.
Whether further merging would be justifiable in order to obtain larger regions and reach a higher signal-to-noise ratio is not a decision to be taken by the algorithm.
If we decided that the differences that are present in the data (e.g. the intensity and shape of the stellar continuum) are not relevant for our purpose of measuring the Balmer lines, we should provide a different input dataset (e.g. a continuum-subtracted spectrum) that is free from such irrelevant information.

When $S/N<3$ (e.g. in the outskirts), Figure~\ref{fig_error_S/N} shows that \batman\ regions are large enough to decrease the errors, even considering the effect of correlated noise.
However, it is precisely in this regime (particularly for the weakest Balmer lines) that \shifu\ tends to fail to identify the emission line and/or return spurious values of both fluxes and errors (e.g. many of the blue and purple dots in the \hg\ and \hd\ error maps).
These `catastrophic' spaxels make a significant contribution to the fraction of spaxels with $S/N<1$.
If they are not merged with the surroundings (as it is often the case when \batman\ considers the `full set' of spectroscopic data), there is little apparent improvement in the signal-to-noise.
If, on the contrary, they are incorporated into the adjacent regions, they may spoil the fit for the whole region, leading to the large blue/purple areas in the `$\pm 15$-\AA' panels for \hg\ and \hd.

Moreover, the \emph{estimated} signal-to-noise $x/e$ is a random variable expected to follow a normal distribution centred at the true value of $S/N$, which has a significant impact on the histograms plotted in  Figure~\ref{fig_error_S/N} whenever the noise is comparable to the signal.
In our example, the colour maps in Figures~\ref{fig_test_galaxy} and~\ref{fig_error_S/N} indicate that both the errors and the amplitude of random fluctuations in the outer regions have decreased by a similar amount when \shifu\ is run on the binned spectra, compared to the spaxel-by-spaxel measurements.
Although this is certainly a valuable improvement, it barely reflects in the distribution of signal-to-noise ratios.

Although it is not the goal of the present work to deal with all these effects, they illustrate the kind of problems that one may face when analysing real scientific data and the way \batman\ handles them by default.
Most importantly, they help showing that the formal errors returned by the algorithm provide an estimate of the true uncertainties roughly as good as the underlying assumptions apply to the particular problem at hand.
Thus, we strongly advise not to blindly take them (nor the estimated signal-to-noise ratios) at face value.

%-------------------------------------------------------------------------
\section{Conclusions}
\label{sec_conclusions}
%-------------------------------------------------------------------------

This article describes the Bayesian Technique for Multi-image Analysis (\batman), a new segmentation algorithm designed to characterize and coherently tessellate simultaneously many layers of a given multi-image, which we define as a dataset containing two regularly-sampled spatial dimensions and an arbitrary number $n_\lambda$ of `spectral' layers, such as e.g. Integral-Field Spectroscopic (IFS) data.

\batman's tessellation attempts to identify spatially-connected regions that are statistically consistent with the underlying signal being constant, given the information (measurements and corresponding errors) provided in the input dataset.
If the difference between any two regions is found to be significant, they are kept separate in order to avoid unnecessary loss of information.
It is important to note that these considerations are independent on the signal-to-noise ratio: if two regions carry the same information (have compatible signal within the errors), they should always be merged together; if they do not, it may be completely unphysical to average over them, and \batman\ will keep them separate.

In order to test the performance of the algorithm and provide some guidance to future users, we have created a set of test cases that comprises both synthetic and real data, analysed in different ways.
According to the results of these tests, we conclude that:
\begin{enumerate}

\item
The output tessellation depends on the precise choice of the input data set, and therefore it is of paramount importance that the user devotes some time to investigate the information that should be considered relevant as a preliminary step of any scientific analysis.

\item
\batman\ adapts to the spatial structures present in the data for a wide variety of morphologies, regardless of the statistical properties of the noise.
By construction, gradients pose a significant challenge to the algorithm.
When they are present, the output tessellation tends to trace the isocontour lines.

\item
The exact number and size of the regions are mainly set by the local signal-to-noise ratio.
The higher the S/N of the data, the easiest it is to distinguish whether two spaxels/regions are different.
When S/N is low, many spaxels may be consistent with having a similar signal, and only a small number of large-size, independent regions can be identified.

\item 
The precise value adopted for the combined prior parameter $K$, the only free parameter of the algorithm, also affects the number and size of the regions in the output tessellation by setting the end of the iterative procedure.
Lower values of $K$ result in more iterations and therefore a smaller number regions.
This may have a substantial impact on the number of (potentially spurious) structures identified on the smallest scales, particularly when $n_\lambda=1$ (`monochromatic' mode).

\item
In the proposed formalism, the expected values \mm\ of the posterior probability distribution of the signal within each region are given by the inverse variance-weighted average~\eqref{eq_mean}.
Our synthetic tests show that these values provide a good representation of the true signal.

\item
The formal errors \sg, given by expression~\eqref{eq_variance}, are indicative of the true uncertainties, but they may underestimate them by as much as a factor of the order of two (e.g. in the presence of gradients and/or spurious regions).

\item
Our analysis of real astronomical data shows that the segmentation of IFS datacubes is a complex task, involving many issues that are specific to the scientific problem under study.
Our tests based on \ngc, focusing on the measurement of several Balmer lines, suggest that \batman\ may be most helpful in the low S/N regime.
The algorithm is capable of recovering the underlying structure of the object even in the most difficult case (\hd), especially when it is applied directly on the CALIFA data (and not so much when the \shifu\ measurements are taken as input).
The reduction of the noise with respect to the spaxel-by-spaxel (no binning) approach is clearly visible in the galaxy outskirts.

\end{enumerate}

As a final remark, let us stress once again that, in contrast to many other segmentation algorithms, \batman\ aims to preserve all the spatial information contained in the original data, as long as it is considered statistically significant.
Such philosophy represents a new (and, in the opinion of the authors, much needed) approach to the analysis of astronomical images in the advent of the vast amount and spatial resolution of IFS data to come.

%-------------------------------------------------------------------------
\section*{Acknowledgments}
%-------------------------------------------------------------------------

Financial support has been provided by research grant AYA2013-47742-C4-3-P from the \emph{Ministerio de Econom\'{i}a y Competitividad} (Mineco, Spain), as well as the exchange schemes `Acciones Conjuntas Hispano-Alemanas' (PPP-Spain-57050803) and `Research Grants - Short-Term Grants, 2016' (57214227) promoted by the \emph{Deutscher Akademischer Austausch Dienst} (DAAD, Germany), and the `Study of Emission-Line Galaxies with Integral-Field Spectroscopy' (SELGIFS, FP7-PEOPLE-2013-IRSES-612701) programme funded by the Research Executive Agency (REA, EU).
YA is also supported by contract RyC-2011-09461 of the \emph{Ram\'{o}n y Cajal} programme (Mineco, Spain).
RGB acknowledges support from grants AyA2014-57490-P and JA-FQM-2828.
This study makes uses of the data provided by the Calar Alto Legacy Integral Field Area (CALIFA) survey (\url{http://califa.caha.es/}), 
based on observations collected at the Centro Astron\'{o}mico Hispano Alem\'{a}n (CAHA) at Calar Alto, operated jointly by the Max-Planck-Institut fűr Astronomie and the Instituto de Astrofísica de Andalucía (CSIC). 
Finally, the authors would like to thank the `Galaxies and Quasars' research group at the Leibniz-Institut für Astrophysik Potsdam (AIP), and Peter Weilbacher in particular, for useful discussions and constructive feedback, as well as the anonymous referee for pointing out several aspects (e.g. Table~\ref{tab_symbols}, Section~\ref{sec_fluctuations}, and Appendix~\ref{sec_stopping}) that have certainly helped to improve the clarity of the paper.

%-------------------------------------------------------------------------

\bibliographystyle{mnras}
\bibliography{references}

%-------------------------------------------------------------------------
\appendix
\section{1D Problem}
\label{sec_1D_problem}

%-------------------------------------------------------------------------

%------------------------------------------------
\begin{figure*}
\centering
\includegraphics[height=.9\textheight]{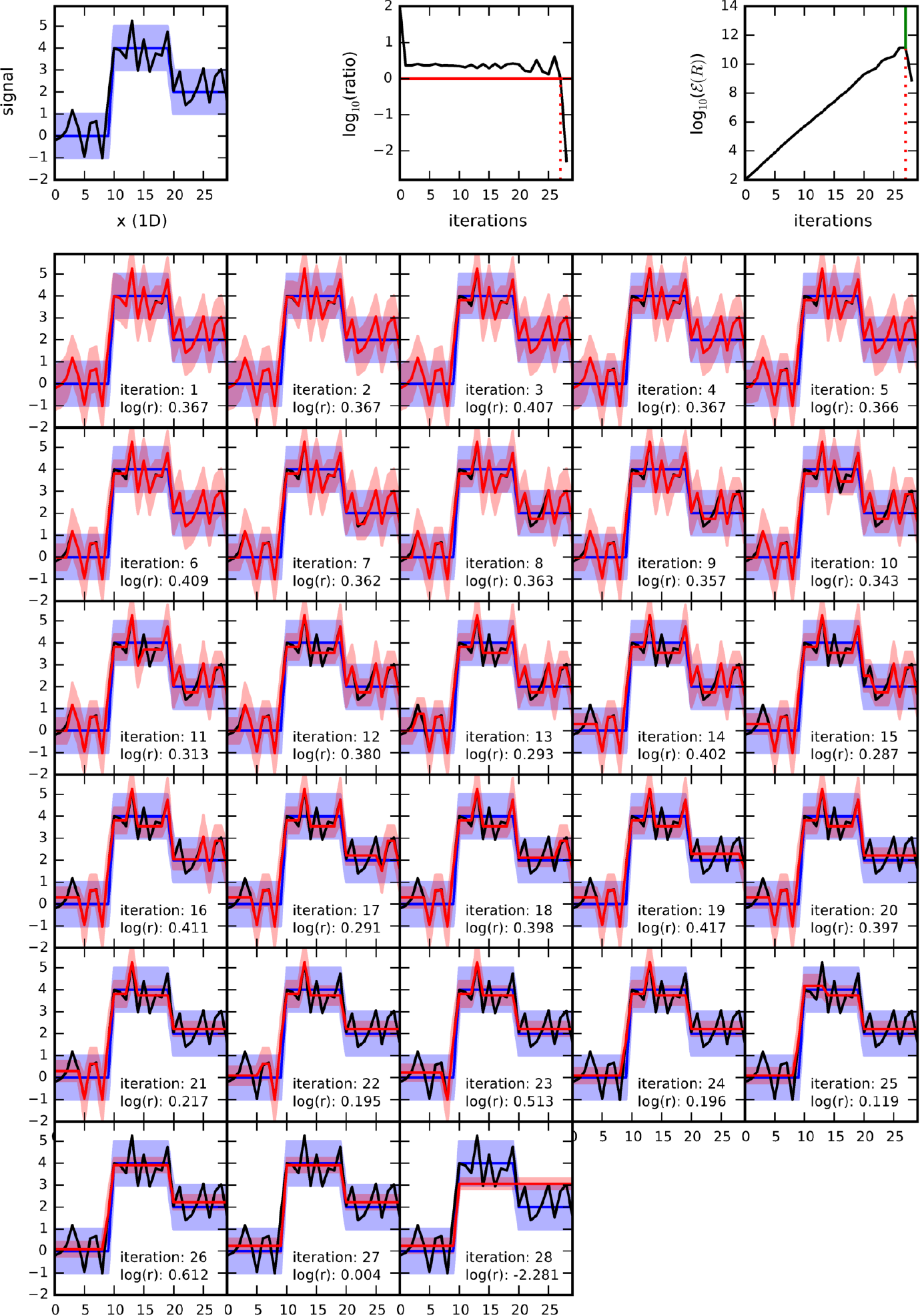}
\caption
{
Illustration of the binning procedure in a 1D problem.
Top left panel shows the true signal (blue line), $1-\sigma$ error (blue shaded region), and a random realization of the measurements (black line).
The value of the ratio~\eqref{eq_posterior_candidate} for the candidate tessellation selected at every iteration and the evidence $\mathcal{E}(R)$ are depicted in the top middle and top right panels, respectively.
The horizontal red solid line on the middle panel indicates the stopping condition (ratio smaller than unity), while the vertical dotted line marks the iteration at which it is attained.
The top right panel shows that it coincides with the maximum value of the evidence (green solid line).
The rest of the panels plot the recovered signal $\mu$ and the estimated error $\sigma$ as a red solid line and red shaded region, respectively.
The logarithm of the ratio (positive until iteration 28) is quoted on every panel.
}
\label{fig_1Dmodel}
\end{figure*}
%------------------------------------------------

Let us illustrate the greedy binning procedure that \batman\ performs in order to explore the space of all possible tessellations by means of a very simple problem in one dimension.
To this purpose, we have created a 30-element vector with values 0, 4, and 2 in the ranges [0,9], [10,19] and [20,29], respectively (blue line in top left panel of Figure~\ref{fig_1Dmodel}.
We have added Gaussian noise with a dispersion $\sigma=1$ (blue shaded region) to the `noiseless' `1D-image', as in the synthetic test cases of Section~\ref{sec_synthetic}, to obtain an input vector (signal and noise) for our binning algorithm (black solid line).

We apply \batman\ to this problem and allow the algorithm to continue binning even when it is no longer `convenient' according to our Bayesian criterion (see Section~\ref{sec_code}), i.e. when the ratio in equation~\eqref{eq_posterior_candidate} is smaller than unity for any new tessellation obtained by merging two adjacent regions, the evidence for any such model decreases with respect to the current solution, and \batman\ considers that any further co-addition of regions leads to a loss of statistically-significant information.

The entire process is presented in Figure~\ref{fig_1Dmodel}, where the top row shows (left) the input (black) and true (blue) signals, as well as the variation of the evidence ratio (middle) and the evidence (right) for the current tessellation $\mathcal{E}(R)$ according to expression~\eqref{eq_evidence_R_full} as a function of the number of iterations (or, equivalently, the number of regions in the tessellation).
The horizontal solid red line (ratio equal to one) in the middle top panel corresponds to the stopping criterion of our algorithm, and the iteration where it is reached is indicated by a vertical dotted red line on both the middle and right panels.
We have verified that this iteration is always close to (and very often coincides) with the absolute maximum of the evidence, marked by the solid green line in the right panel (see Appendix~\ref{sec_stopping}).

The rest of the panels (second to seventh rows) show the iterative merging of vector elements into larger one-dimensional regions.
In addition to the true signal (blue) and input measurements (black) we have plotted the recovered signal (red solid line) and estimated error (red shaded area) at every iteration.
\batman\ starts by merging the most similar vector elements, which makes iterations 1 (connecting elements 16 and 17) and 2 (elements 10 and 11) almost unnoticeable (the underlying black curve is barely visible).
However, one can see that the noise (red shade) is reduced (see iteration 2) even when the change in the recovered signal is very small.
The binning process continues, and the red line progressively resembles the blue one, while the red shade tightens around it (the noise is reduced).
Iteration 27 shows the most likely tessellation (last one with a positive logarithm of the evidence ratio), with iteration 28 (last panel) containing only two regions being clearly less representative of the input signal, not only in mathematical terms but also from visual inspection.

~\\

%-------------------------------------------------------------------------

\section{Stopping criterion}
\label{sec_stopping}

%-------------------------------------------------------------------------

% %------------------------------------------------
\begin{figure*}
\centering
\includegraphics[height=.3\textheight]{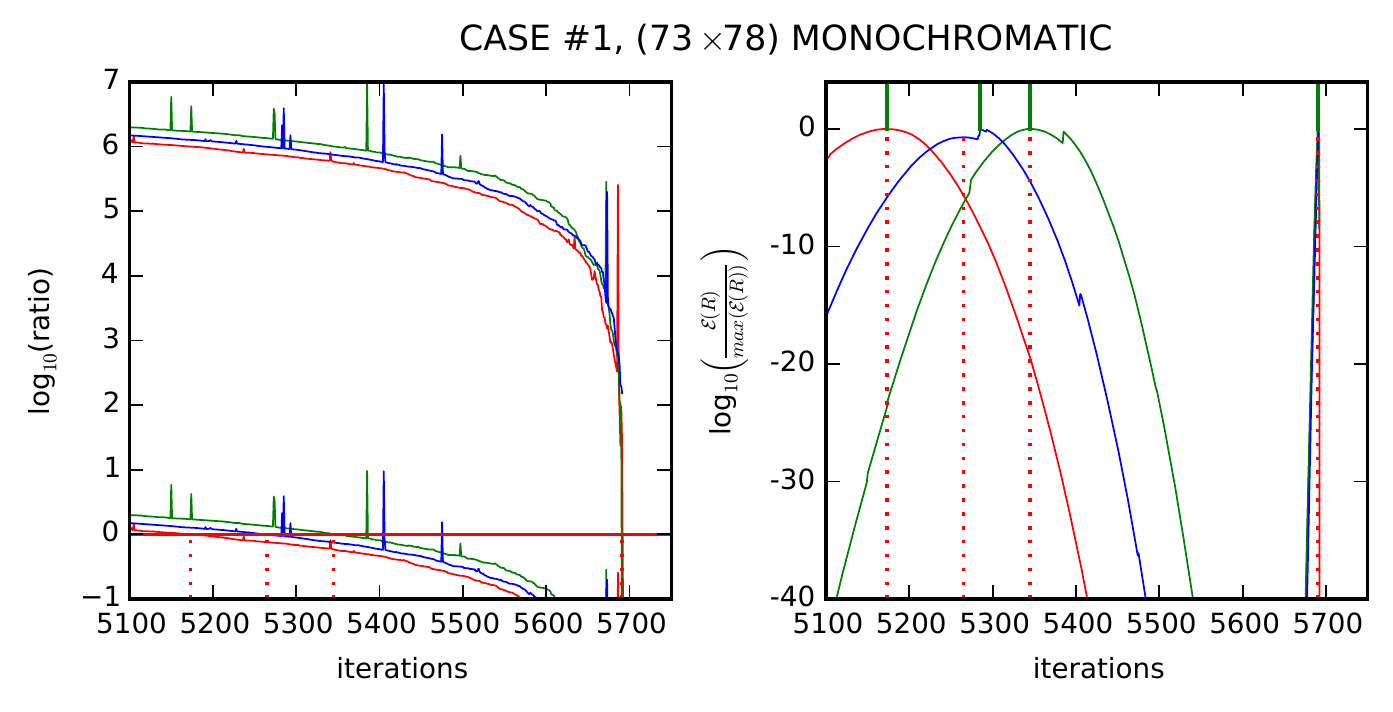}\\
\includegraphics[height=.3\textheight]{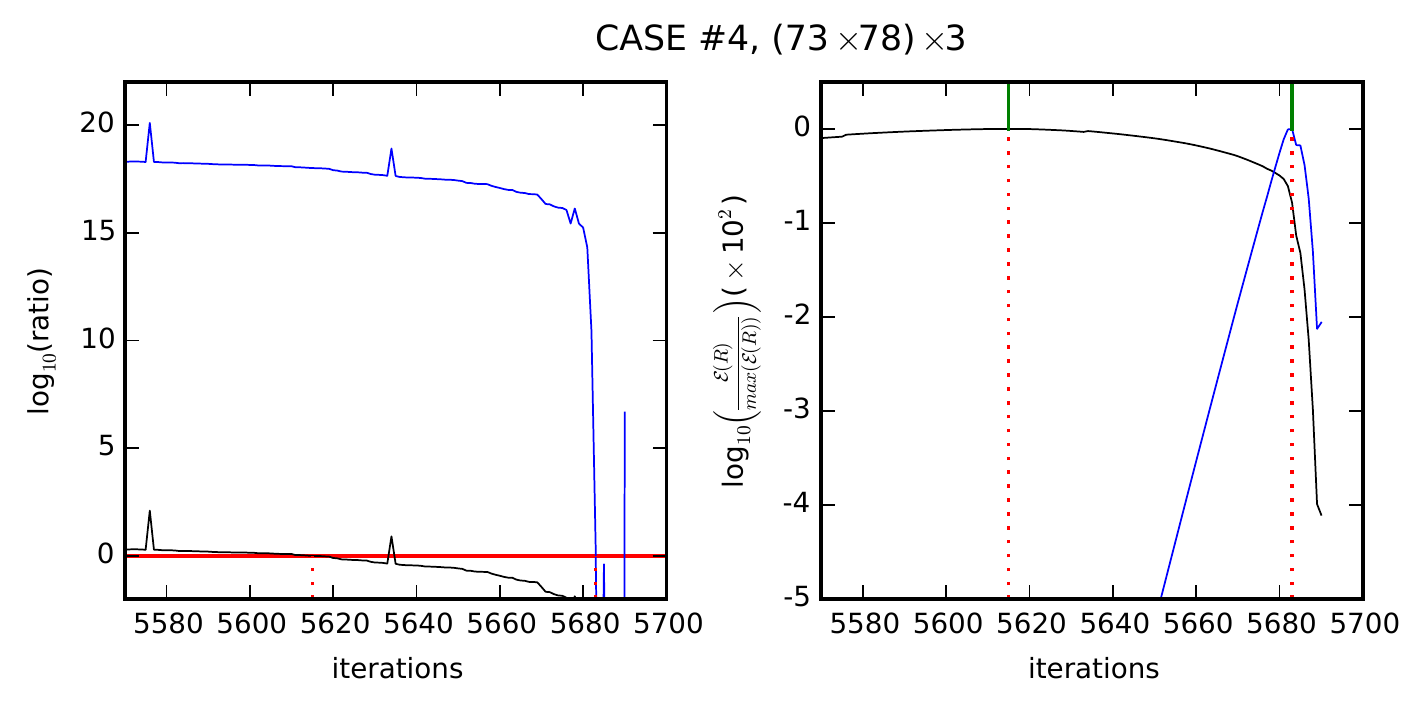}\\
\includegraphics[height=.3\textheight]{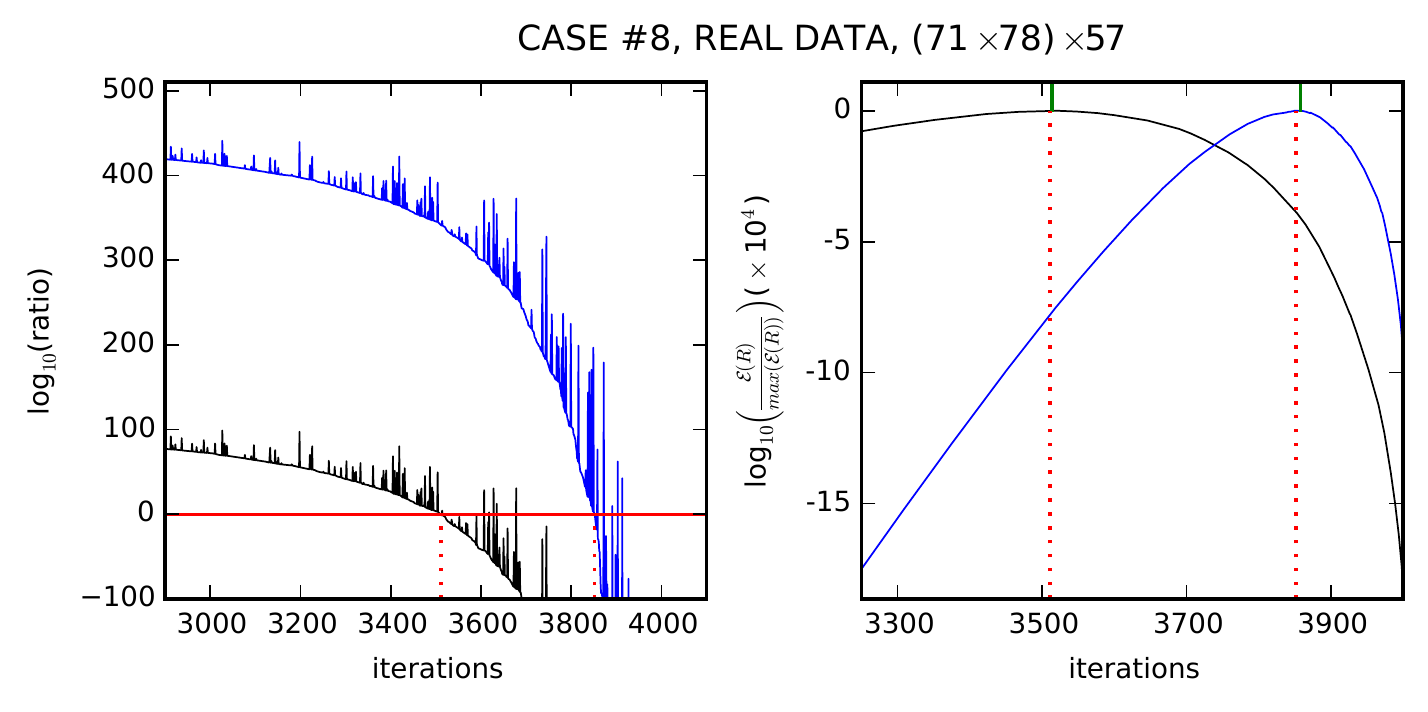}\hspace*{7mm}
\caption
{
Illustration of the stopping criterion (red solid and dotted lines) in relation to the ratio (left) and the evidence (normalized for comparison, right) at every iteration for the test cases \#1 (top), \#4 (middle), and \#8 (bottom). Top panels present 6 curves corresponding to the two different values of the K parameter (1 lower curves and $10^{-6}$ higher ones) for the different RGB images. Middle and bottom panels show two curves corresponding to the different values of the parameter K (1 in black and $10^{-6}$ in blue). Red solid line (left panels) represents the threshold used as stopping condition. Red dotted line (both panels) marks the stopping iteration. Green solid line (right panels) corresponds to the maximum of the evidence. Values for the stopping and evidence maximum iterations are depicted in Table~\ref{tab_it}.
}
\label{fig_local_maxima}
\end{figure*}

In principle, there is no guarantee that our greedy procedure does not converge to a local maximum of the evidence rather than finding the global optimum.
While more elaborate (e.g. Monte Carlo) techniques would allow a more exhaustive exploration of the full tessellation space, here we present the results of a simpler test to show that the adopted stopping criterion, i.e. evidence ratio~\eqref{eq_posterior_candidate} smaller than unity, yields a nearly optimal segmentation, even when it does not always coincide with the absolute global maximum of the Bayesian evidence $\mathcal{E}(R)$.

As we did for the one-dimensional problem presented in Appendix~\ref{sec_1D_problem}, we have re-run \batman\ for our test cases \#1 (monochromatic mode on R, G, and B images with uniform noise, i.e. 3 times), \#4 (multi-$\lambda$ mode, RGB image with CCD-like noise) and \#8 (real data, full set, directly binning the IFS datacube) allowing the algorithm to proceed beyond the adopted stopping condition.
Figure~\ref{fig_local_maxima} shows the ratio (left column) and the evidence (right column) as a function of iteration number for each of these test cases.
Once again, the red solid horizontal line and the red dotted vertical line indicates \batman's stopping criterion.
In most cases, the evidence does not increase further if we allow for additional merging of regions (as shown in Figure~\ref{fig_1Dmodel} for the one-dimensional test problem discussed in Appendix~\ref{sec_1D_problem}), and the stopping condition coincides almost exactly with the maximum evidence, marked by a vertical green line on the right column.

These results suggest that our condition based on the evidence ratio provides a valid (and efficient) criterion to select the final tessellation.
Even in those cases where the stopping iteration does not coincide with the maximum of the evidence (e.g. cases \#1 and \# 8), the values are reasonably close (see Table~\ref{tab_it}).
Nonetheless, we would like to point out once again that, as can be readily seen in Figure~\ref{fig_local_maxima}, the Bayesian evidence does depend significantly on the adopted value of the combined prior parameter and the chosen input data set.
These (problem-dependent) decisions, especially the latter, are much more relevant than the stopping criterion in determining the optimal segmentation returned by the algorithm.
The importance of experimenting by trial and error in each particular case can hardly be overstressed.

%------------------------------------------------
\begin{table}
\begin{center}
\begin{tabular}{llcc}
\hline
test case & K & stopping it. & max($\mathcal{E}(R)$) it. \\
\hline
\# 1 - R & 1         &  $5173$   &    $5173$  \\
         & 10$^{-6}$ &  $5691$   &    $5691$  \\
\# 1 - G & 1         &  $5345$   &    $5345$  \\
         & 10$^{-6}$ &  $5690$   &    $5690$  \\
\# 1 - B & 1         &  $5265$   &    $5285$  \\
         & 10$^{-6}$ &  $5691$   &    $5691$  \\
\hline
\# 4     & 1         &  $5615$   &    $5615$  \\
         & 10$^{-6}$ &  $5683$   &    $5683$  \\
\hline
\# 8     & 1         &  $3511$   &    $3514$  \\
         & 10$^{-6}$ &  $3852$   &    $3858$  \\
\hline
\end{tabular}
\end{center}
\caption
{
Values of the stopping (third column) and maximum evidence (fourth column) iterations, corresponding to the red dotted and green solid lines in Figure~\ref{fig_local_maxima}.
}
\label{tab_it}
\end{table}
%------------------------------------------------

%-------------------------------------------------------------------------

\label{lastpage}
\end{document}